\documentclass[preprint,times,tighten]{aastex631}
\usepackage{xcolor}
\usepackage{multirow}
\usepackage[normalem]{ulem}
\usepackage{enumitem}
\usepackage{amsmath}
\usepackage{cleveref}

\received{Jan 9, 2022}
\accepted{July 12, 2022}
\submitjournal{The Astrophysical Journal}

\shorttitle{Thermal and Non-Thermal properties or AR Jets}
\shortauthors{Paraschiv, Donea, \& Judge}

\begin{document}

\title{Thermal and Non-Thermal Properties of Active Region Recurrent Coronal Jets}

\correspondingauthor{Alin Razvan Paraschiv}

\author[0000-0002-3491-1983]{Alin R. Paraschiv}
\affiliation{High Altitude Observatory, National Center for Atmospheric Research, PO Box 3000, Boulder CO~80307-3000, USA}
\affiliation{School of Mathematical Sciences, Monash University, 9 Rainforest Walk, Clayton, Victoria 3800, Australia}

\author[0000-0002-4111-3496]{Alina C.  Donea}
\affiliation{School of Mathematical Sciences, Monash University, 9 Rainforest Walk, Clayton, Victoria 3800, Australia}

\author[0000-0001-5174-0568]{Philip G. Judge }
\affiliation{High Altitude Observatory, National Center for Atmospheric Research, PO Box 3000, Boulder CO~80307-3000, USA}

\begin{abstract}

We present observations of recurrent active region coronal jets and derive their thermal and non-thermal properties, by studying the physical properties of the plasma simultaneously at the base footpoint, and along the outflow of jets. The sample of analyzed solar jets were observed by SDO-AIA in Extreme Ultraviolet and by RHESSI in the X-Ray domain.
 
The main thermal plasma physical parameters: temperature, density, energy flux contributions, etc. are calculated using multiple inversion techniques to obtain the differential emission measure from extreme-ultraviolet filtergrams. The underlying models are assessed, and their limitations and applicability are scrutinized. Complementarily, we perform source reconstruction and spectral analysis of higher energy X-Ray observations to further assess the thermal structure and identify non-thermal plasma emission properties.

We discuss a peculiar penumbral magnetic reconnection site, which we previously identified as a ``Coronal Geyser''. Evidence supporting cool and hot thermal emission, and non-thermal emission, is presented for a subset of geyser jets. These active region jets are found to be energetically stronger than their polar counterparts, but we find their potential influence on heliospheric energetics and dynamics to be limited. We scrutinize whether the geyser does fit the non-thermal erupting microflare picture, finding that our observations at peak flaring times can only be explained by a combination of thermal and non-thermal emission models. This analysis of geysers provides new information and observational constraints applicable to theoretical modeling of solar jets.
\end{abstract}
\keywords{Solar physics (1476), Solar corona (1483), Solar active regions (1974), Solar extreme ultraviolet emission (1493), Solar x-ray emission (1536), Solar magnetic reconnection (1504), Astronomy data analysis (1858)}

\section{Introduction}
This work aims to concurrently estimate the thermal and non-thermal properties of Active Region (AR) coronal jets and geysers by comparing the assumptions and results of different plasma inversion methods, provide constraints for coronal modeling, and discuss their significance in the context of larger-scale coronal manifestations. 

Coronal jets are observed in ultraviolet, X-Ray, and coronagraphs and are described as reconnection driven ubiquitous small-scale collimated plasma eruptions that are ejected towards the outer corona, presumably into the interplanetary medium where they might provide mass supply to the solar wind flux \citep{stcyr1997}. The \citet{savcheva2007} observational study of small-scale polar jets in the X-Ray domain was fundamental in understanding physical and thermal components of coronal jets, measuring outflow parameters of a significant sample of eruptions. A more recent comprehensive review on solar coronal jets is presented by \citet{raouafi2016}.

AR jets have only recently became a very active research topic due to the improvement of spectroscopic and imaging instrumentation in radio, extreme ultraviolet (EUV), and X-Ray domains. The magnetic configuration associated with AR jets tends to be more complex, making these eruptions hotter and larger when compared to polar jets \citep{moore2010,sako2013}. 68\% of solar X-Ray jets, found by \citet{shimojo1996} and \citet{shimojo2000}, originated in or near active regions, and were associated to micro/nano class flares. In this work, we aim to discuss the energetic classification of jets, when including non-thermal contributions.  

 \citet{nistico2009, nistico2011} provided observational evidence in favor of ubiquitous small-scale reconnection as an intrinsic feature of the solar corona. Typical plasma physical characteristics of polar jets were described and lower-bound electron temperatures for jet outflows between $0.8-1.3 \cdot 10^6$ K were realistically estimated via Differential Emission Measure (DEM) inversions. Enhanced results for polar jets were subsequently obtained using higher quality data or improved inversion schemes \citep[see ][]{pucci2013,young2014,paraschiv2015}. \citet{sterling2015}, proposes flux cancellations as the main driver of coronal jets, and \citet{mulay2016,mulay2017} present parameter estimates in favor of flux cancellation as the main driver for recurring AR jets, using detailed filtergram and spectroscopic observations. The bulk outflow of erupting material was estimated at $T_e\sim 2\cdot 10^6$ K and plasma number density $n_e\sim 1\cdot 10^{10}$ cm$^{-3}$, with energy inputs appearing substantially higher when compared to polar jet estimates. The authors differentiated a secondary temperature peak, $T_e\sim 4-5\cdot 10^6$ K, which they attributed to the flaring footpoint. We aim to discern if such observed higher temperature components are due to instrumental and/or inversion effects or if indeed AR jets can be, at least in our sample case, multi-thermal.
 
The `hot' and `cool' terms regarding eruptions are loosely used in the literature based on the discriminant observations that are pursued. For example, \citet{mulay2017b} define cooler eruptions as being in the $\log T_e/K \sim 5$ range, while other works define cool eruptions manifesting in the $\log T_e/K\, 3-4$ range \citep{canfield1996}. Similar particularities affect the `hot' attribute. Caution is needed when comparing results from different studies. We adopt the convention: typical coronal DEM temperatures are between $\log T_e/K\, 6.0-7.0$,  'hot' corresponding to temperatures $\ge\log T_e\, 7.0$ K and 'cool' to $\le\log T_e\, 6.0$ K.
 
A morphological dichotomy of coronal jets was proposed based on observational features and associated emission mechanisms \citep[see ][Fig. 1 and Fig. 10]{moore2011}. According to this scheme, standard jets follow a classic x-type reconnection picture, usually associated with flux emergence, while blowout jets involve more impulsive reconnection in a more complex topology, usually associated with small filament eruptions. Expanding the study, \citet{moore2013} showed significant similarities between standard and blowout jets. Concurrently, AR jets were simulated in 3D by \citet{moreno2008} by introducing a sheared flux rope in a tilted magnetic field configuration. The system initially formed a current sheet, which in turn reconnected releasing jet-like eruptions, reaching electron temperatures in the range of $\log T_e/K \sim 7$. The magnetic topology revealed a spire configuration, similar to blowout observations. The simulations of \citet{archontis2010} showed that jets can take place in short successive recurrent phases. Analyzing jets originating in solar coronal holes, \citet{moreno2013} modeled a set of recurring jets that resemble mini-CMEs suggesting that their properties may resemble blowout jets. Further observational evidence provided by \citet{sterling2015} suggested that the blowout scenario involving mini-filaments may be responsible for both classified event types with no fundamental physical difference. Furthermore, \citet{muglach2021} showed that flux emergence and rotational drivers are insufficient to explain their jets.

\begin{figure}[!t]
\hspace{-0.4cm}\includegraphics[width=1.04\linewidth]{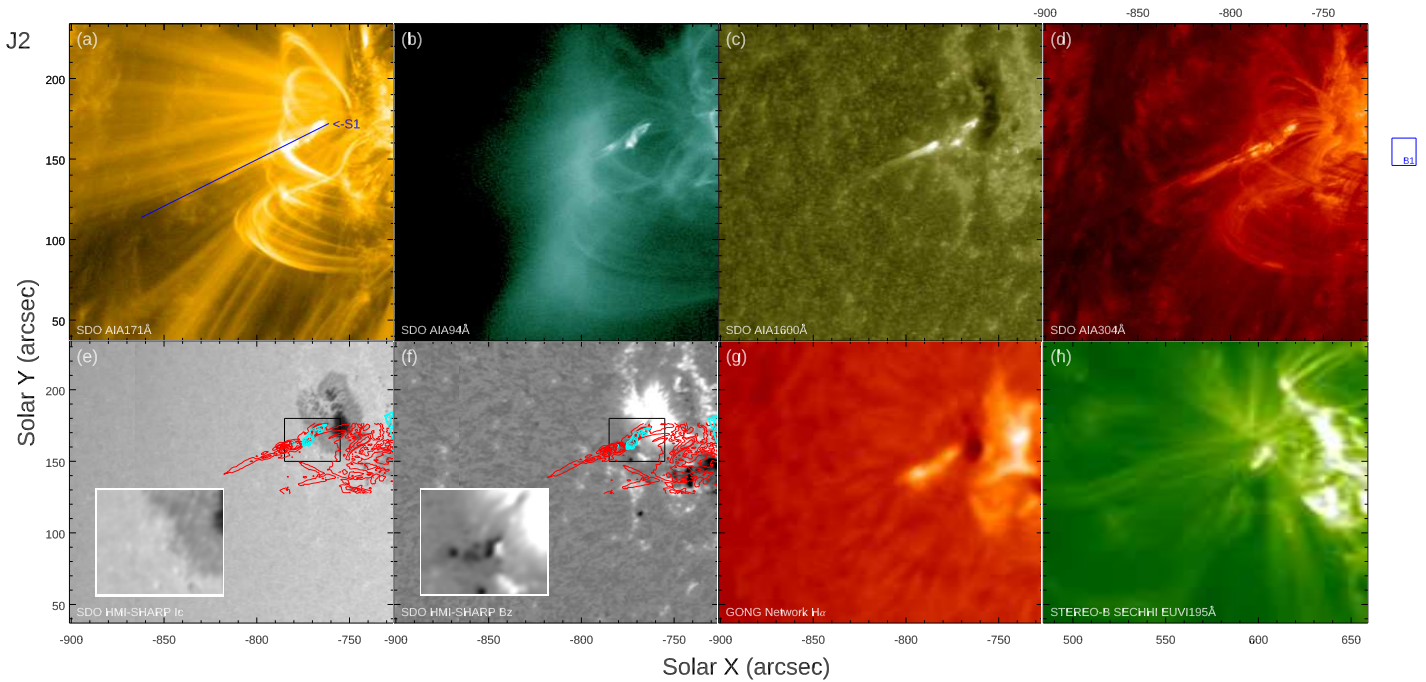}
  \vspace{-0.7cm}\caption{Coronal jet J2,  25.09.2011 at 01:13:09UT as revealed by multiple observations. The figure sub-panels represent different correlated manifestations of jet activity: (a) SDO AIA-171\AA{} and timeseries tracked slit (S1) position; (b) AIA-94\AA{}; (c) AIA-1600\AA{}; (d) AIA-304\AA{}; (e) SDO HMI intensity continuum with AIA-304\AA{} (red) and AIA-94\AA{} (blue) contours, (f) HMI SHARP (active region patch) vertical $B_z$ magnetogram with AIA-304\AA{} (red) and AIA-94\AA{} (blue) contours; (g) BBSO H$\alpha$; (h) STEREO-B EUVI-195\AA{} different viewpoint observation. }
\label{fig-alldata}
\end{figure} 

The current literature describes fast recurring jets, usually originating from one site during short periods of time, typically 2-8 hours \citep[e.g.][]{innes2011,schmieder2013,chen2015,sterling2016,hu2016,liu2016,chen2021}. Longer observations are presented by \citet{panesar2016} and \citet{paraschiv2019}. Fast recurring jet observations were first discussed by \citet{chifor2008}. They found a correlation between recurring magnetic flux cancellation close to a pore and the X-Ray jets. Using EUV data, \citet{guo2013} and \citet{schmieder2013} discussed fast recurring AR coronal jets aiming to understand their morphology, reporting twisting motions, and inconsistencies between observations and existent models. \citet{chen2021} examine recurrent jets, their reconnection sites, and their role in triggering CME's.  The behavior of such sites on longer timescales and detailed emission mechanisms leaves room for further exploration.

This work addresses `recurrence' by studying a unique long-lived flaring site, also known as the jet's brightpoint or more generally as a flaring footpoint, that successively generated jets. Our recurrent jets are presumably generated via small-scale flaring events that occur at the same spatial location.  Due to the long timescale (24 h) of our studied recurring jets we have assumed the existence of quasi-stable erupting structures, and following earlier nomenclature \citep{Menzel2013} we define \textbf{Coronal Geysers} \citep{paraschiv2018,paraschiv2019,paraschiv2020} as long-lived small-scale penumbral AR structures that have an open field connectivity with roots in complex magnetic configurations, that are subject to helicity conservation and can contain filamentary structures. Geysers are prolific at generating recurrent jets, radio bursts and energetic particles, and are classified in this work as impulsive microflare sites.

In this work, we evaluate the thermal and non-thermal emission of three coronal jets that originated from one such geyser, calculate their energy budget and discuss the implications to larger scale coronal phenomena. Both EUV and X-Ray observations and methods are described in sec. \ref{sec-obs}, while Sec. \ref{sec-aia} describes the EUV observational results for both the main jet outflow and footpoint. The methods used to invert the DEM from the EUV data, the applicability to our data and the main results are discussed. Next we showcase the results obtained from analyzing X-Ray sources and spectral fitting of thermal and non-thermal emission components (Sec. \ref{sec-rhessi}). In Sec. \ref{sec-concl}, we discuss the DEM results and debates the jet energetics and implications in a coronal context. Furthermore, evidence for downwards acceleration of particles is presented and compared to microflare statistics. For completeness, a short summary of community-driven DEM inversion methods used herein, along with their application and limitations with respect to small-scale jets is presented separately in Appendix \ref{sec-appendix}.

\section{Observations, Methods, and Instrumentation}\label{sec-obs}

\subsection{The penumbral AR11302 Geyser}\label{sec-obs:sub-obs}

We previously highlighted a recurrent jet site that was detected at the SE periphery of AR11302 on 25 Sep. 2011. Continuous observations spanning 24 h revealed 10 EUV jets, labeled J1-J10, erupting from an unique footpoint, the geyser \citep{paraschiv2019,paraschiv2020}.  

 From a thermal perspective, \cref{fig-alldata} shows multi-wavelength observations of one AR jet in the EUV and ultraviolet channels of the  Solar Dynamics Observatory, \citep[SDO;][]{pesnell2012} Atmospheric Imaging Assembly \citep[AIA;][]{lemen2012} and the Solar Terrestrial Relations Observatory \citep[STEREO;][]{kaiser2008}, Extreme Ultraviolet Imager \citep[EUVI;][]{wuelser2004}. The AIA and EUVI instruments show the geyser's activity from different vantage points. All jets followed the same propagation direction.

 The SDO-AIA sub-panels (\cref{fig-alldata} a-d) show multi-wavelength observations of the J2 jet. The figures corresponding to all 10 jets can be found in \citet{paraschiv2018}. Sub-panels a-d show emission in the AIA-171\AA{} and AIA-94\AA{} coronal filters along with the AIA-1600\AA{} and AIA-304\AA{} transition region and chromospheric filters. The filters sample a wide range of plasmas that erupt simultaneously. Clear morphological differences can be observed. A slit S1 (shown in \cref{fig-alldata} a) was selected to correspond to the footpoint location of all observed recurrent jets and follows the jets outflow. 

Additional SDO Helioseismic and Magnetic Imager \citep[HMI; ][]{scherrer2012} SHARP \citep{bobra2014} and BBSO H$\alpha$ context data are presented. \Cref{fig-alldata} e-g reveal the lower atmosphere structures involved in generating jets. Contours of hot AIA-94\AA{} (blue) and cool AIA-304\AA{} (red) plasma are over-plotted to pinpoint the location of the jets. In a separate work \citep{paraschiv2020}, we assess potential jet trigger mechanisms through measurements of vector magnetic fields in the lower solar atmosphere.

This work focuses on a subset of three jets. The non-thermal eruption components could be scrutinized using observations from the Reuven Ramaty High-Energy Solar Spectroscopic Imager \citep[RHESSI;][]{lin2002} only in the cases of J2, J3, and J6. The other 7 jets were not observed by RHESSI. More details are presented in Sec. \ref{sec-obs:sub-rhessi}. 

\subsection{SDO-AIA Methodology}\label{sec-obs:sub-sdo}

SDO-AIA provides full-disk solar images, observing the Sun in 7 EUV, 2 UV, and 1 white light channels, with a spatial platescale resolution of $\sim 0\farcs 6$ pix$^{-1}$ and temporal cadence of 12 s. The SDO data was obtained using the JSOC pipeline\footnote{\href{http://jsoc.stanford.edu/ajax/exportdata.html}{http://jsoc.stanford.edu/ajax/exportdata.html}} and processed to level 1.5 using standard and custom implementations of calibration procedures; e.g. coalignment, respiking, aia\_prep corrections, exposure normalization, etc. All data was preprocessed using the Solarsoft (SSWIDL) package\footnote{\href{http://www.lmsal.com/solarsoft/}{http://www.lmsal.com/solarsoft/}}. We have respiked all the SDO-AIA data in preparation for the DEM analysis, as smaller dynamic features can be misidentified by the AIA despiking algorithm \citep{young2021}.

\begin{figure}[!t]
  \begin{center}
\includegraphics[width=1.0\linewidth]{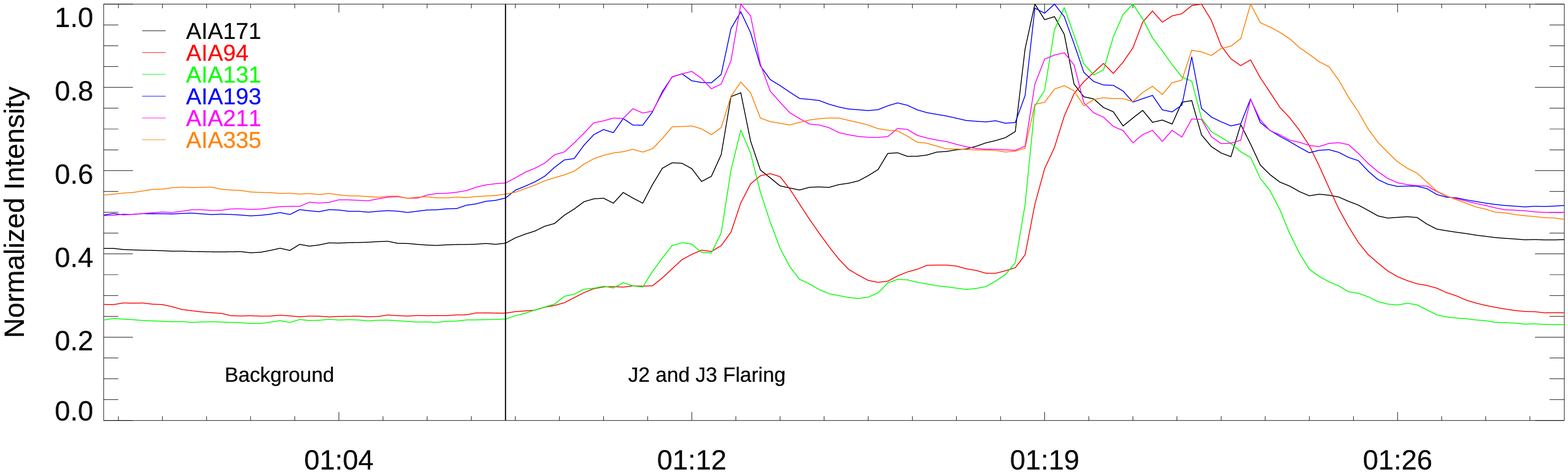}
\includegraphics[width=1.0\linewidth]{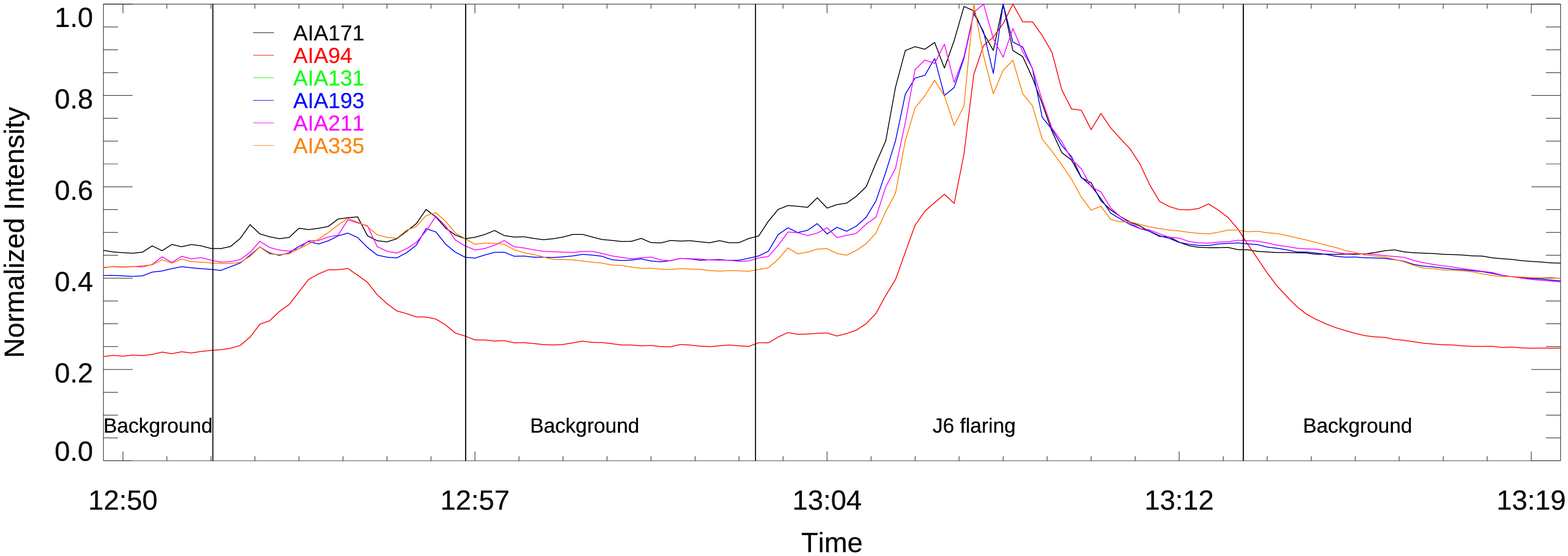}   
  \end{center}
 \caption{Normalized intensity timeseries of the six utilized SDO-AIA filters, temporally centered around the three footpoint eruptions. The intensity was averaged inside the fixed individual footpoint borders, illustrated by the contours in the AIA-131\AA{} sub-panel of \cref{fig-jfil}. The top panel corresponds to the temporally close J2 and J3 while the bottom panel corresponds to the J6 jet. Vertical lines separate temporal slots representative of typical background intensities.}
\label{fig-jitot}
\end{figure}   

The EUV thermal geyser component is analyzed using the SDO-AIA data. The filtergrams are characterized by a multi-thermal emission line contributions over a broad temperature range. Six SDO-AIA filter centered on EUV bandpasses  [94\AA{}, 131\AA{}, 171\AA{}, 193\AA{}, 211\AA{}, 335\AA{}] are used for DEM inversions. These filtergrams are centered on iron emission lines (e.g. \ion{Fe}{8}, \ion{Fe}{9}, \ion{Fe}{12}, \ion{Fe}{14}, \ion{Fe}{16}, \ion{Fe}{18}, etc.) theoretically sampling plasma in the $T_e=[0.4,30]$ MK range. The AIA-304\AA{} filter is not suitable for DEM analysis \citep{warren2005}. 

 Observations of plasma formed at higher temperatures, such as Hinode  \citep{kosugi2007} XRT \citep{golub2007}, RHESSI, or FOXSI \citep{krucker2013} based observations can in principle be used to add information from plasma in the higher temperature range. Multiple studies \citep{cheung2015,hanneman2014,inglis2014,mulay2017b,athiray2020} performed joint inversions of EUV and X-Ray imaging data, demonstrating that, when available, adding X-Ray data to EUV filtergrams can greatly improve the accuracy of coronal plasma determinations. Only RHESSI observations were available for this geyser.
 
Here, DEM inversions were performed using multiple inversion models where method assumptions, when applied to jets, are compared, and output total Emission Measure (EM) are cross-validated. The consolidated inversion outputs are then used to discuss physical implications of jet eruptions. The total EMs as opposed to the more commonly used DEMs were used by us in order to clearly reveal the total amounts of electron plasma, seen at a specific temperature, where we divided the logarithmic temperature range into linear bins of $T_{bin}=\log T_{e}$. Additional details are presented in app. \ref{sec-appendix:sec-theory}.

The simplest DEM interpretation, namely the filter ratio technique (see app. \ref{sec-appendix:sub-fr}) was initially attempted, but showed to not be appropriate for SDO-AIA data \citep{paraschiv2018}.  The \citet{aschwanden2013} (henceforth `A2013', see app. \ref{sec-appendix:sub-teem}) method is a simple and straightforward implementation of a single Gaussian fitting solution optimized using a $\chi^2$ minimization. The \citet{hannah2012} (henceforth `H2012', see app. \ref{sec-appendix:sub-demreg}) approach optimizes an initial `guess' solution obtained by a similar $\chi^2$ minimization via $SVD$ solutions for filtergram ratios fitted inside bins spaced along an empirical temperature range. The \citet{cheung2015b} inversion (henceforth `C2015`, see app. \ref{sec-appendix:sub-sparse}) uses the Simplex algorithm to reduce an under-determined linear system, where the two independent variables are the evenly spaced temperature bins and the number of available filtergrams.  
  
  Comparing the results from the A2013, H2012, and C2015 should theoretically yield similar EMs, though differences will exist based on the assumptions implicit to each method. The A2013 method described in \ref{sec-appendix:sub-teem} provides two quantities for the plasma inside the LOS volume: the temperature integrated DEM and the DEM-weighted average temperature. In order to recover an integral EM for comparison to the C2015 and H2012 results we employ the approximation given by eq. \ref{sec-appendix:eq-demem}. Our application of the H2012 method utilizes eq. \ref{sec-appendix:eq-dem2em} in order to convert from DEM to EM values. Each EM is correspondent to its $\log T_{e}\pm \sigma_{T_e}$ temperature space. Where required, EMs are then transformed to plasma number density $n_e$ using eq. \ref{sec-appendix:eq-neapprox} assuming a $\phi=1$ filling factor. We did not consider SDO-AIA responses for $\log T_e/K>7.3$. This is an already optimistic assumption given the flat nature of SDO-AIA response at these temperatures. 

 We have developed scripts, calibrations, and adaptations of the methods and publicly available inversion codes described above. Additionally, the SDO-AIA response curves were customized to our observational parameters, namely: we applied temporal orbital degradation (via the $timedepend\_ date$ keyword); normalized intensities using the SDO-EVE full disk measurements; used the CHIANTI \citep[V8.0x;][]{delzanna2015} atomic database. These corrections substantially alter the output results when compared to the default configurations of the inversions. 

\subsection{RHESSI Methodology}\label{sec-obs:sub-rhessi}

RHESSI was a NASA small explorer mission, operating between 2002 and 2018, and investigated the X and $\gamma$ ray EM, energetics, and particle acceleration of solar flares. RHESSI records spectroscopic data using nine rotating collimator grids, in front of a spectrograph, covering the entire solar disk. RHESSI can perform high energy imaging of hot solar features via Fourier transform analysis of timeseries data from its rotating collimators with a spatial resolution as fine as 2". The timeseries is inverted to yield an emission map, assuming that the sources do not change during integration. We have used 7 our of 9 detectors for this study. Detectors 2F and 7F were dismissed as they exhibit flat responses and problems with photon calibrations for the duration of our observations.

A general overview of the RHESSI imaging is provided by \citet{hurford2002}. Different source reconstruction techniques are available as part of the RHESSI data analysis SolarSoft package, SSW/HESSI. The pixon solution is quoted to be the most accurate in terms of spatial domain and photon distribution \citep{hurford2002}.  As the geyser dataset requires just a few reconstructions where the best accuracy possible is desired, we utilized the Pixon method to reconstruct the geyser morphology. A pixon represents the abstraction of a pixel-like cell structure, where the information (e.g. number, size, width, etc.) in such a cell depends on a measured global quantity. 

The imaging reconstruction capabilities are not the only strong-point of RHESSI. The satellite's primary data product is the X-Ray spectrometric flux measurements. The geyser's energy spectrum can be extracted, calibrated, and fitted against a vast repository of thermal and non-thermal analytical functions. The RHESSI spectral analysis was performed usingthe GUI SSW/OSPEX module. 

The RHESSI X-Ray source location could be reconstructed and the spectral analysis could be performed for the three geyser jet footpoints. We reiterate that the other seven jets were not observed by RHESSI. In the complementary set of seven, two distinct issues hindered analysis: (i) The lightcurves corresponding to a subset of geyser jets were being masked by stronger AR flaring, occurring in temporal proximity; (ii) Events occurred during RHESSI data gaps. 

\begin{figure}[!p]
\begin{center}
\includegraphics[width=1.01\linewidth]{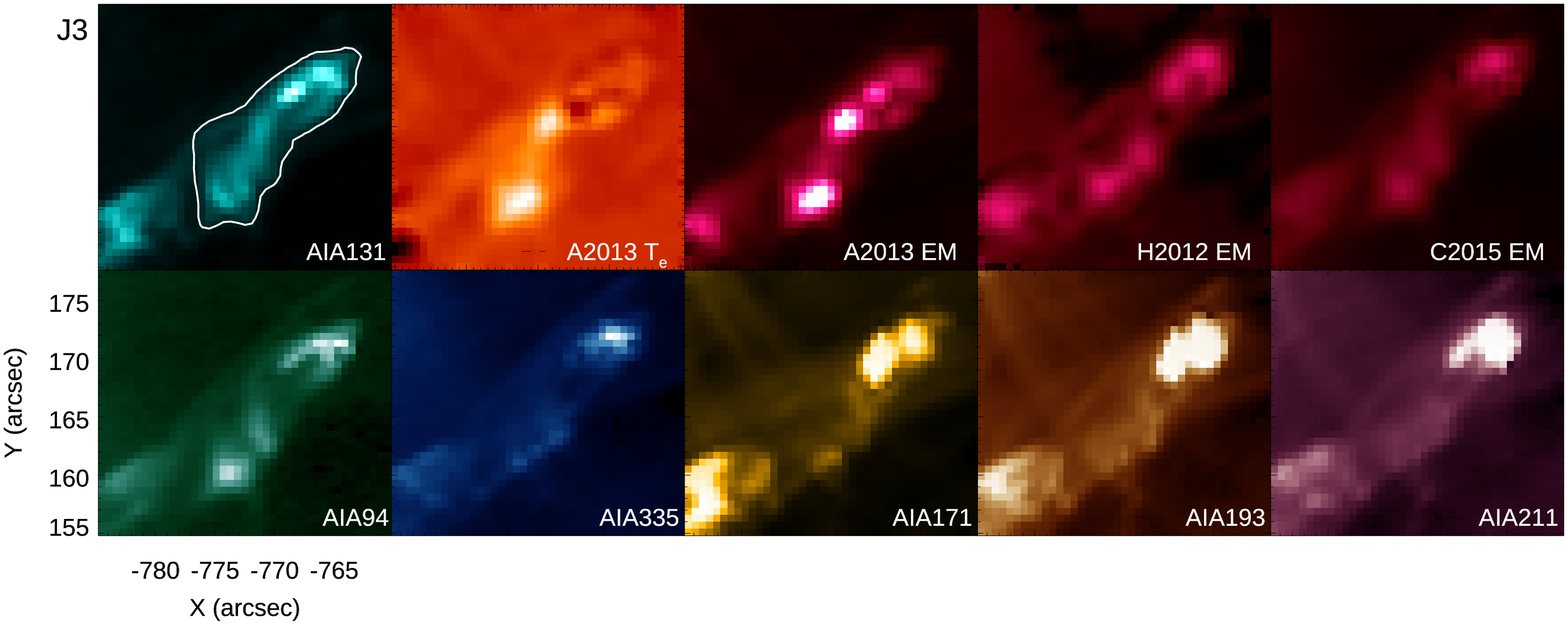}

\includegraphics[width=1.01\linewidth]{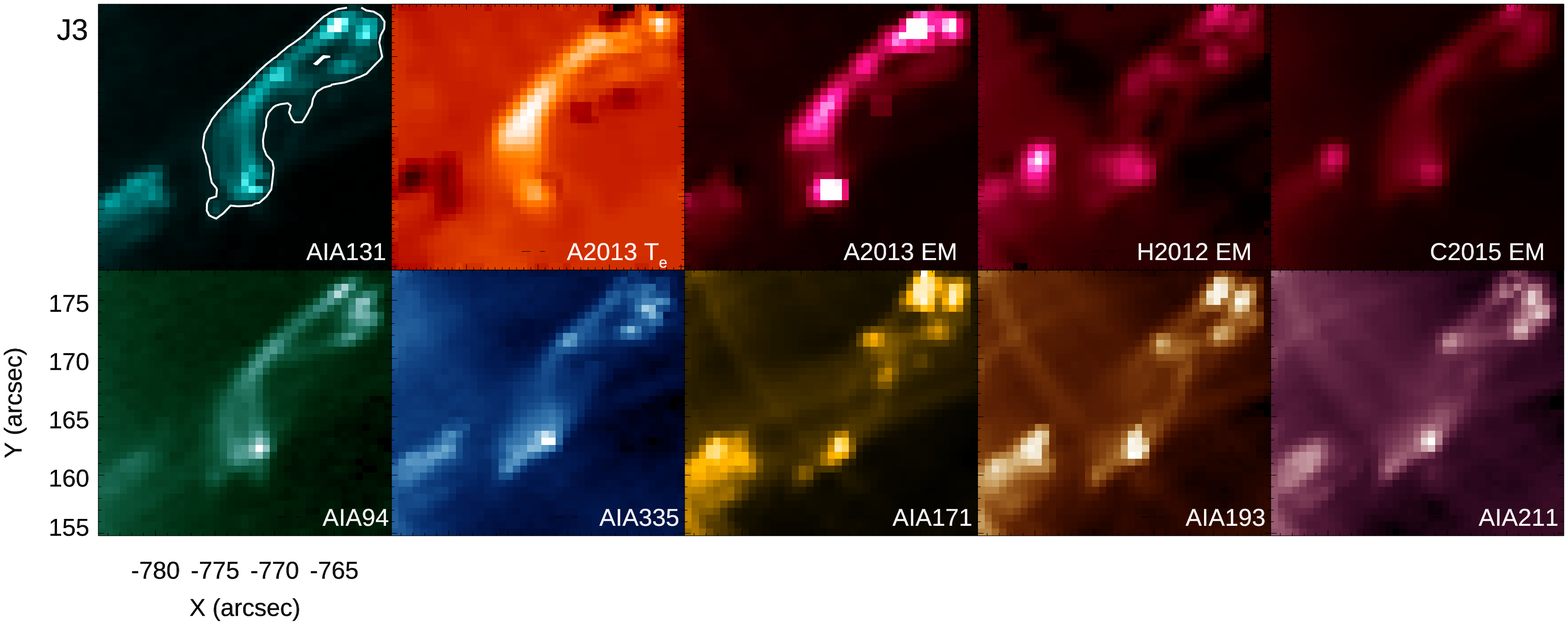}

\includegraphics[width=1.01\linewidth]{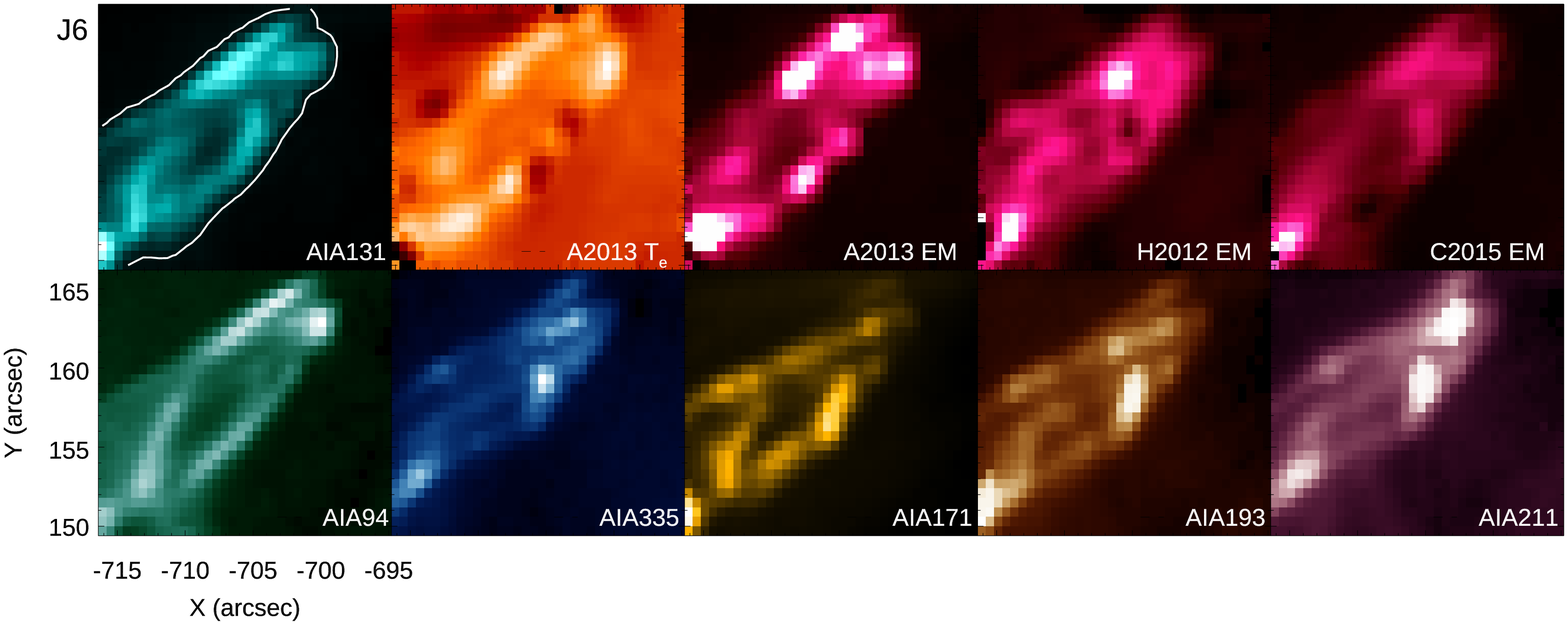} 
\end{center}
\caption{Qualitative observational features of the J2, J3, and J6 footpoints. The Coronal Geyser is revealed for each SDO filtergram, DEM inversion, and peak temperature maps around maximum flaring times: 01:13:00UT for J2 (Top), 01:19:00UT for J3 (Middle), and 13:08:24UT for J6 (Bottom) respectively. Each of the depicted AIA filters exhibits a unique morphology, where the AIA-131\AA{} and AIA-94\AA{} filters best resemble the DEM emission maps. The total EM maps are integrated  between $5\cdot 10^{27} - 5\cdot 10^{28}$ cm$^{-5}$. The A2013 $T_e$ map is integrated inside the $\log T_{e}/K=[5.7,7.0]$ interval.  The contours corresponding to AIA-131\AA{} represent the distinct borders of the observed flaring footpoints, at peak emission time, as corresponding to each jet.}
\label{fig-jfil}
\end{figure}

\section{Analysis and Results (I): The Geyser EUV Thermal Emission} \label{sec-aia}

\subsection{The Geyser Footpoint Emission Measures}\label{sec-aia:sub-jfoot}

The SDO-AIA filtergram background and flaring intensities for all three jet footpoints are presented in \cref{fig-jitot}. The timeseries of SDO-AIA counts are averaged inside fixed jet footpoint regions illustrated as contours in the AIA-131\AA{} panels of \cref{fig-jfil}. Temporal intervals describing both pre-flare background and flaring intensities are highlighted.

The J2 and J3 jets background corresponds to observations between 01:00-01:08UT. In the case of J6 three data subsets without footpoint activity sampled the background intensity. These are: 12:50UT-12:52UT, 12:57UT-13:03UT, and 13:13UT-13:19UT. The J2, J3 and J6 flaring periods are 01:09UT-01:14UT, 01:18UT-01:24UT, and 13:06UT-13:12UT, respectively. The AIA-94\AA{} filter reveals a consistent small time delay of $12-24$s (data cadence is $12$s) for each eruption onset, followed by a slower cooling phase when compared with the other filtergrams. We note that our footpoints exhibit only very few saturated pixels in small localized patches, usually under 10\% of the total region, which were excluded from analysis and quantitative estimations.

In order to accurately recover the footpoint EM, some geometrical approximations are required. In \cref{fig-jfil},  the AIA-94\AA{} and AIA-131\AA{} filters reveal two main flare loops. These filters are adequate in representing hot loop morphology due to the large contribution to the response function from high temperature plasma. The other filters mainly sample lower temperatures.

The flaring loops can not be easily spatially separated. As the observation is not LOS disentangled, these measurements are influenced by projection effects, governed by the inclination with respect to the local vertical of the sun. The width $(W)$ and height $(H)$ of the flaring loops are estimated and summarized in \cref{table-allparam}. The total width does not represent the sum of the individual widths as the features are superposed, but is obtained as an average width across the length, including the areas where the two loops appear separated. The separation area was included in the width estimation as it exhibited systematic stronger emission than the background along all SDO-AIA filters. The flaring loops had a typical diameter of $2467 \pm 436$ km (or 6$\pm1$ pix.), irrespective of the individual jets. Assuming a cylindrical geometry, this loop width can be used to approximate the depth of the emitting region.

\begin{figure}[!t]
  \begin{center}
\includegraphics[width=0.495\linewidth]{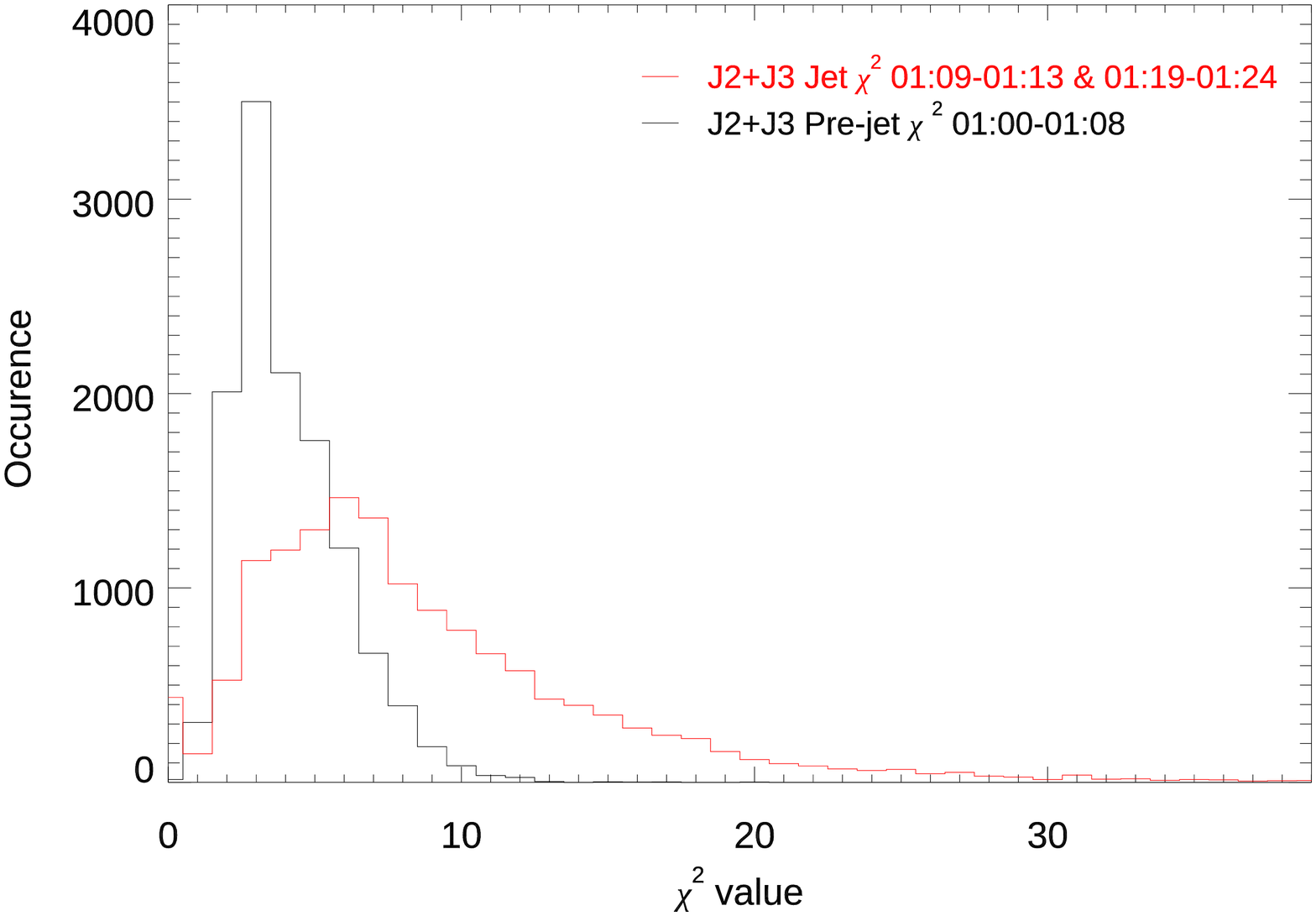}
\includegraphics[width=0.495\linewidth]{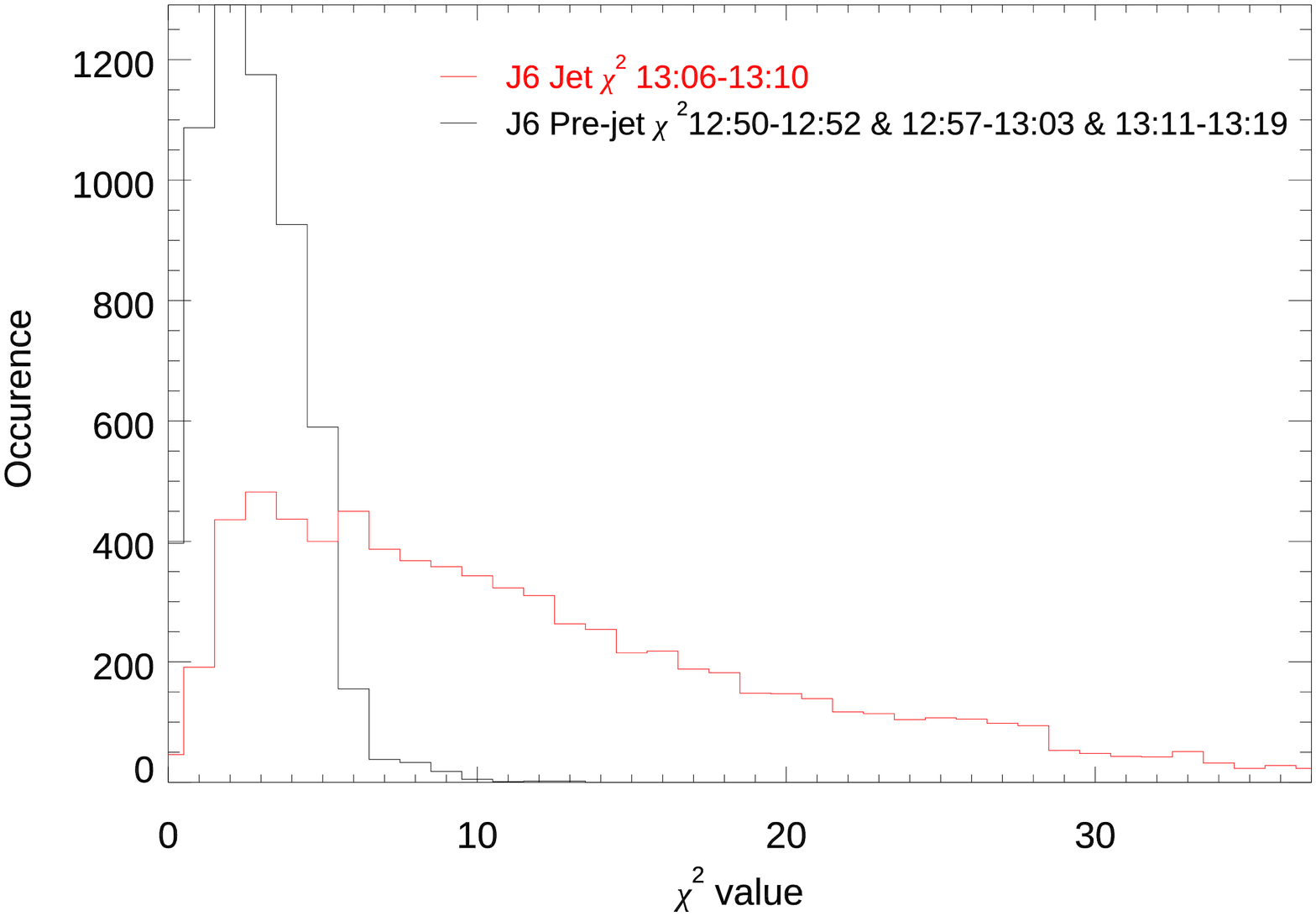}

\includegraphics[width=0.495\linewidth]{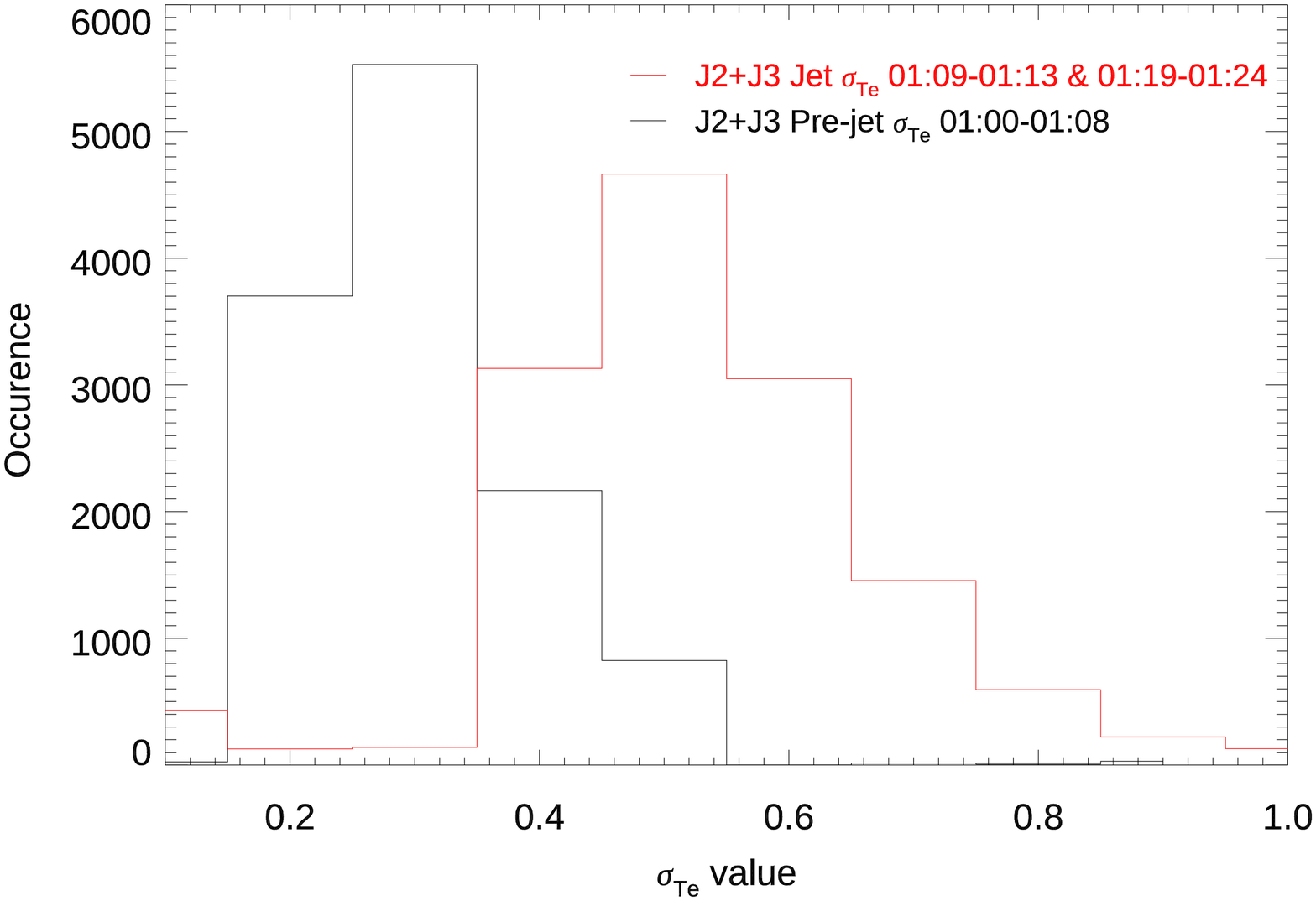} 
\includegraphics[width=0.495\linewidth]{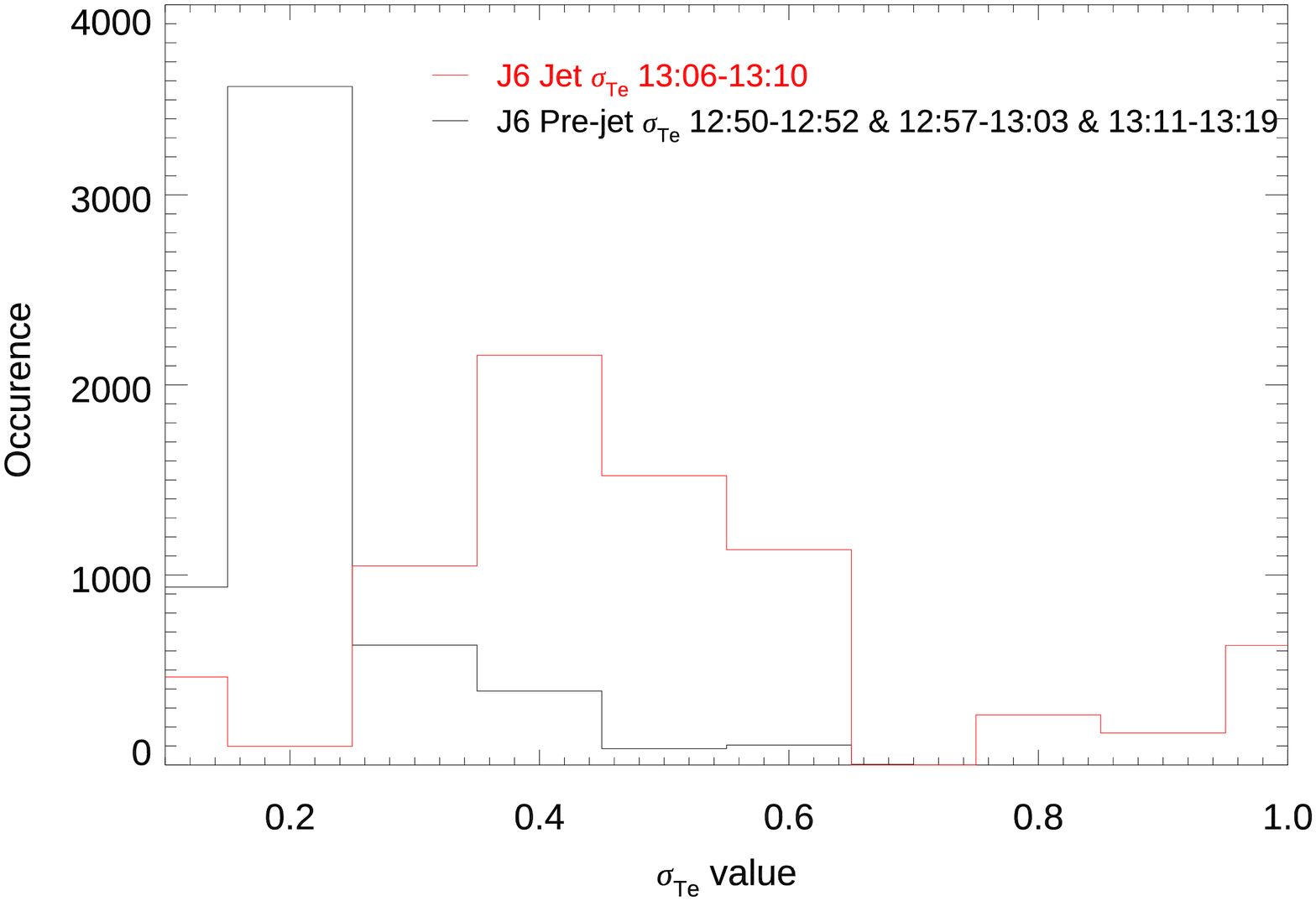} 
  \end{center}
\caption{SDO-AIA A2013 inversion uncertainty distributions inside the selected jet emitting region for pre and post flare conditions (black) and during flaring times, where the goodness of fit degrades (red). Top Panel: $\chi^2$ fit goodness for background conditions are compared to peak flaring times. Bottom Panel: The temperature widths $\sigma_{T_e}$ comparing background and peak flaring conditions. The distributions are comprised from the individual fitting results of the pixels corresponding to the footpoint area and the temporal interval corresponding to flaring and background conditions. }
\label{fig-jfootchi}
\end{figure}

A practical consideration on the validity of inversion methods for this dataset is provided in \cref{fig-jfootchi}. The errors corresponding to the A2013 method present an expected picture of $\chi^2$ based inversion results. For a generic coronal plasma in pre or post jet conditions(black), the fitting residuals are small $(<4)$ indicating a well constrained solution, if following the estimates found in the literature \citep[e.g.][]{aschwanden2011}, but not-so-much from a statistical point of view. We find that accuracy is lost during peak flaring times, even when dealing with these small-scale eruptions (red). This effect occurs due to the saturation in some filters which leads to non-physical responses. Additionally, the uncertainty given by the maximum temperature width $\sigma_{T_e}$ detected pixel-wise inside the flaring time interval is also significantly larger when compared to quiet sun conditions. The J2 and J3 $\chi^2$ residuals, even in quiet conditions are higher than the ones depicted for J6. The issue most probably originates from projection effects and hot loops from above the geyser that are not completely removed by background subtraction due to their dynamics. The H2012 $\chi^2$ metrics exhibit an analogous behavior.

\begin{figure}[!t]
\begin{center}
\includegraphics[width=0.494\linewidth]{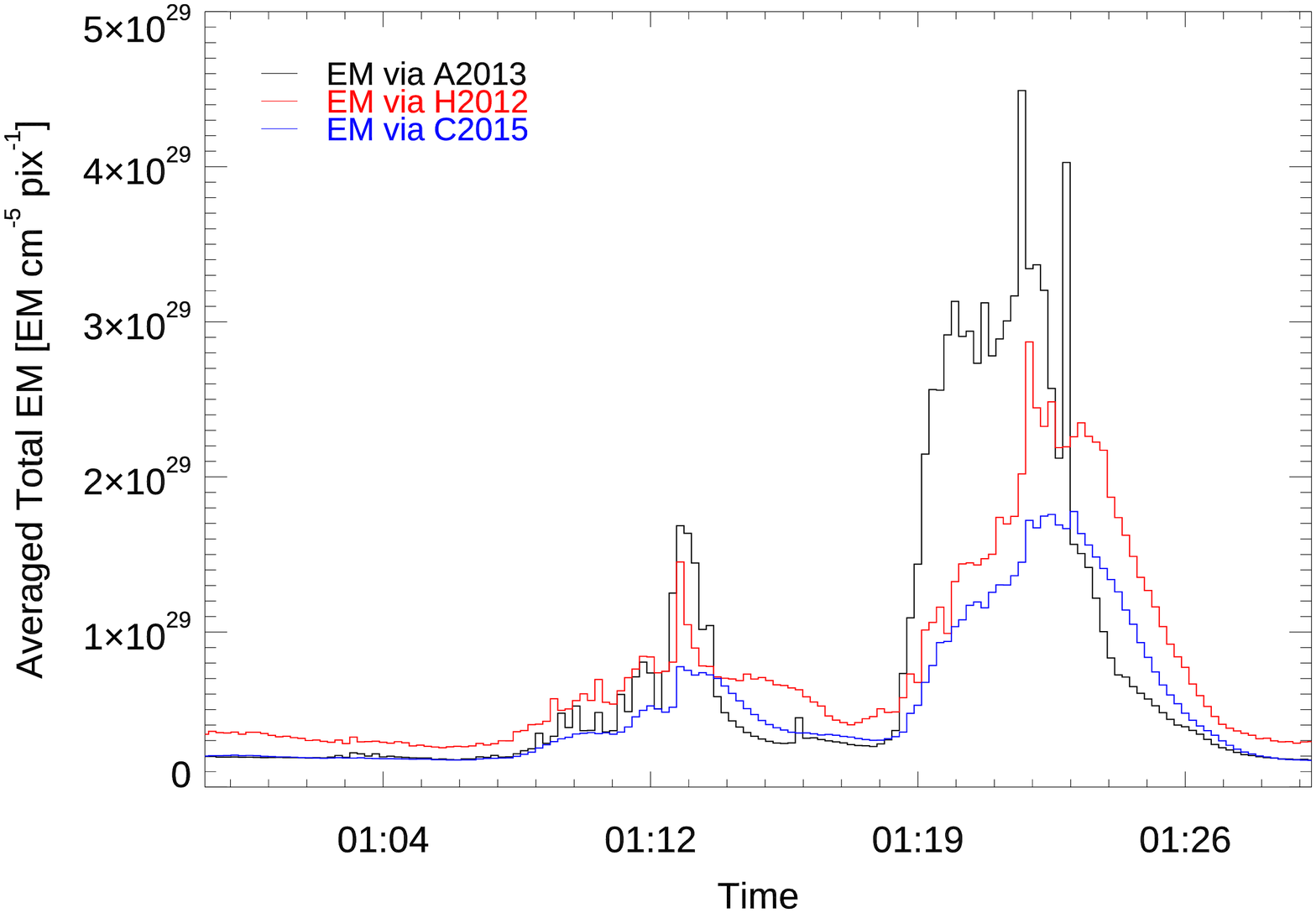}
\includegraphics[width=0.494\linewidth]{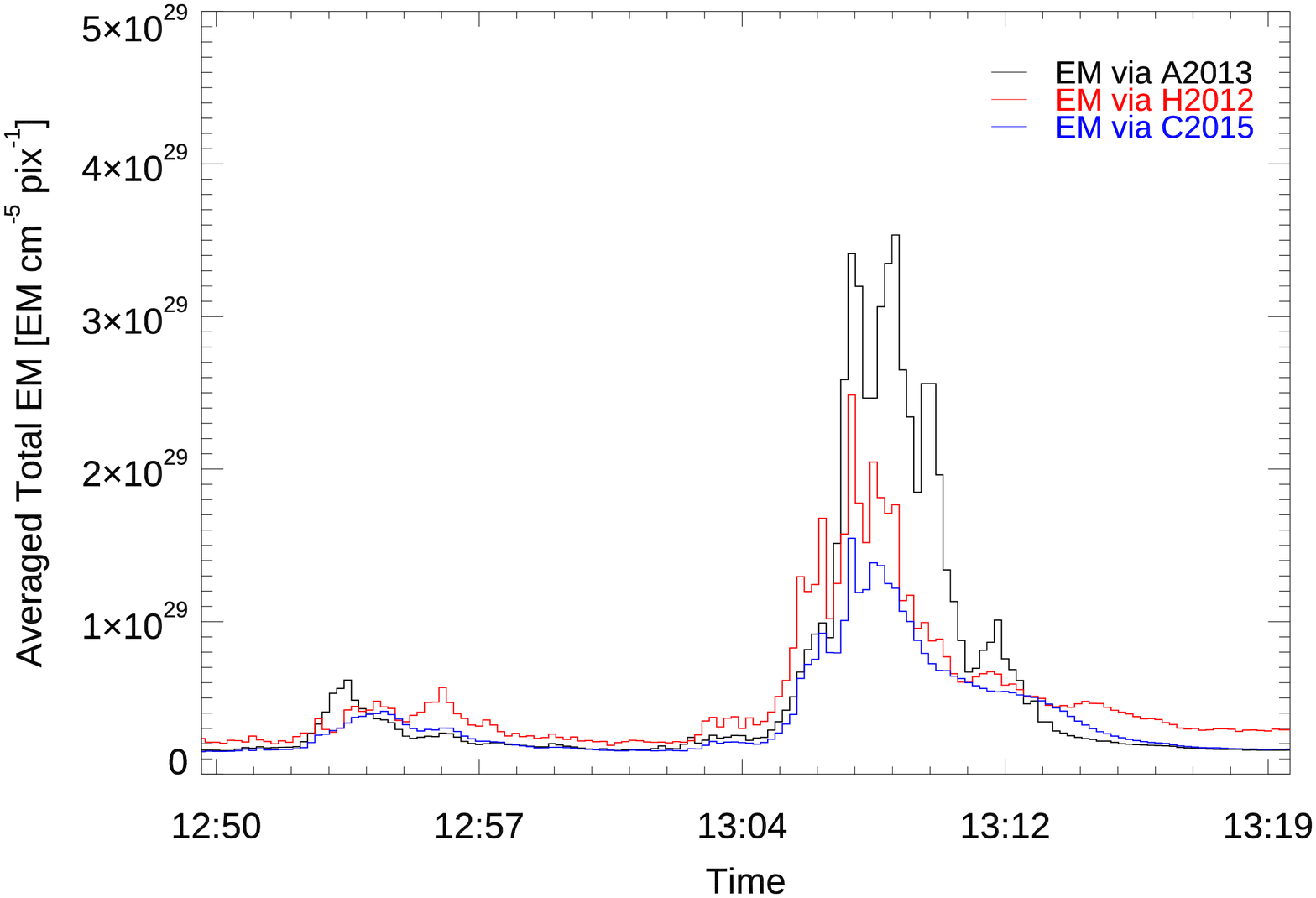}

\includegraphics[width=0.494\linewidth]{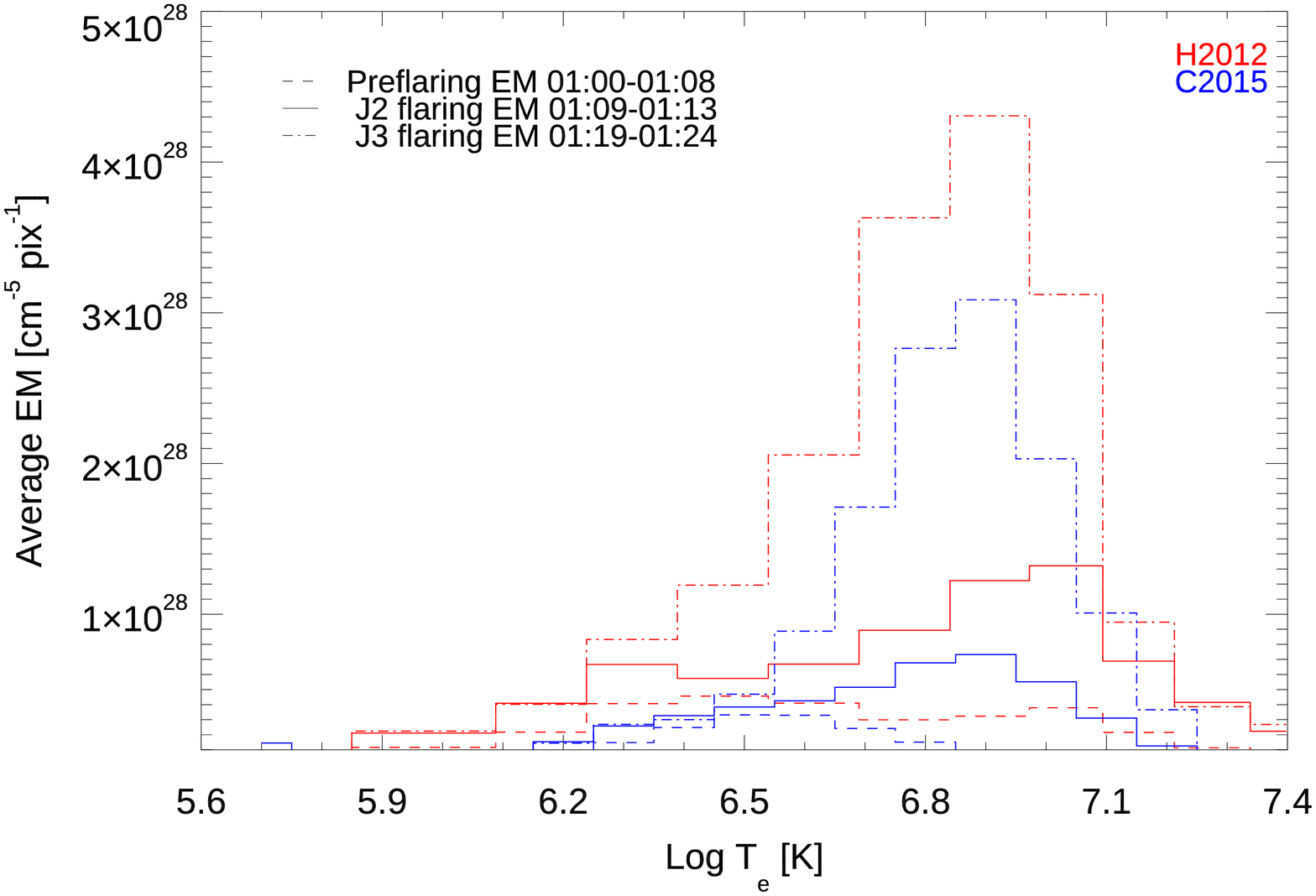} 
\includegraphics[width=0.494\linewidth]{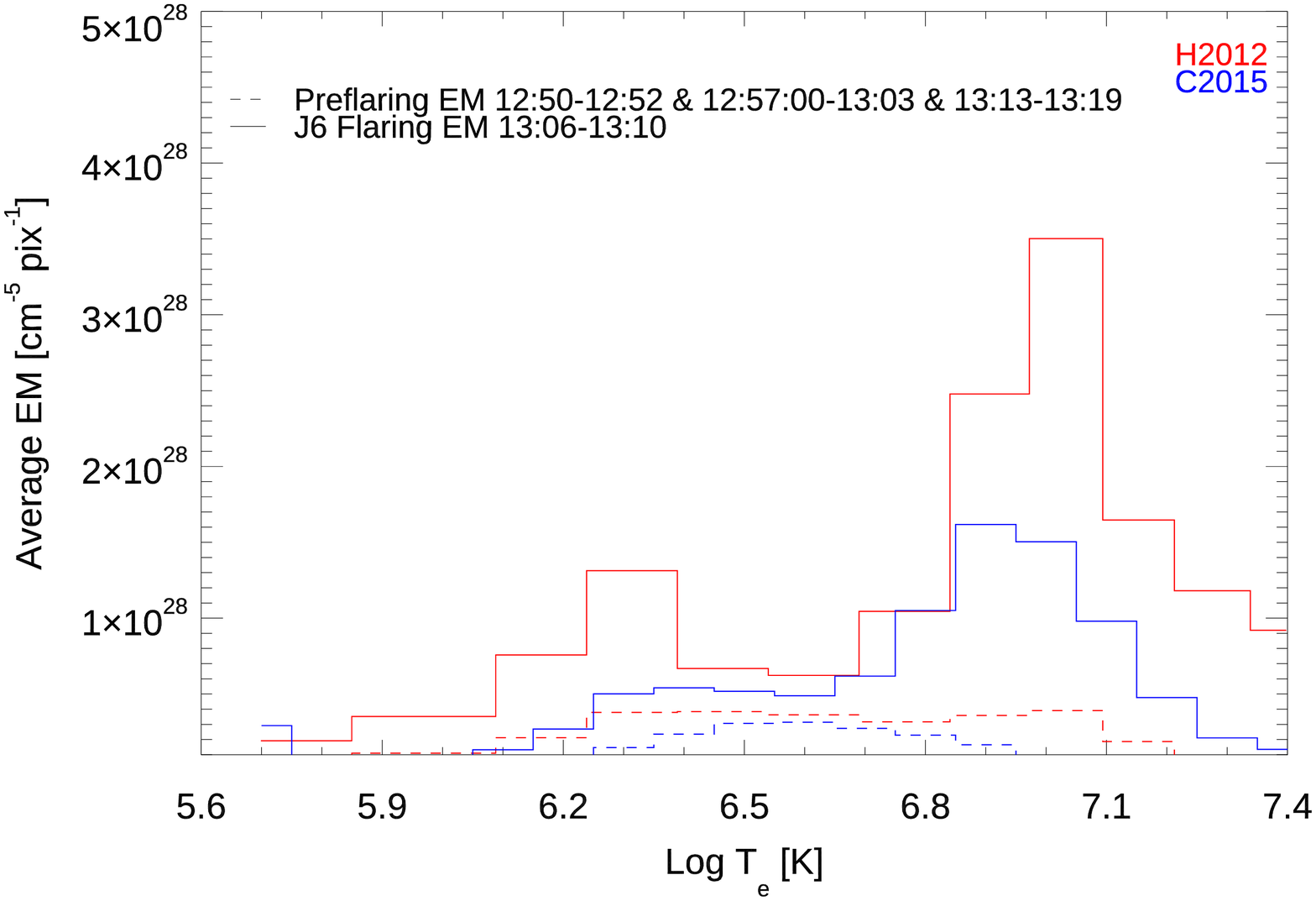} 
  \end{center}
\caption{SDO-AIA EM results corresponding to the three flaring events  recovered via the three inversion methods. EMs are pixel averaged across the footpoint areas. Top Panel: Total average EM over the dominant $T_{e}\pm\sigma_{T_e}$ range for A2013, and over the $\log T_e/K \sim [5.7,7.3]$ temperature range recovered from the H2012 and C2015 inversion methods. A qualitative agreement occurs between the different approaches, although small systematic differences can be observed. Bottom Panel: Temporally averaged EM plotted against the electron $T_{e}$ bins for the selected temporal intervals of quiet and flaring conditions via H2012 and C2015 inversions. Each eruption has a unique temperature distribution.}
\label{fig-jfootem}
\end{figure}

\Cref{fig-jfootem} (top) compares the total temperature integrated EM recovered using the three EM inversions. The EMs are integrated inside a temperature range of $\log T_e/K = [5.7,7.3]$.  The background coronal emission appears negligible when compared to flaring periods.  The A2013 method will fit a region that may well be multi-thermal due to the LOS projection of the main jet emitting material above the flaring site. We found the C2015 EM to consistently return lower counts. We note that the C2015 and H2012 inversion solutions allow for non-zero EM in high temperatures bins (e.g. $\log T_e > 7.0$). However, we reiterate that we did not consider emission from plasma at  $\log T_e/K > 7.3$. This emphasizes that the total EM results are slightly under-determined. Nonetheless, all three eruptions are qualitatively well constrained EM wise inside these temperature ranges.

The total EM plotted in \cref{fig-jfootem} (top) can be further refined for each jet by assessing the shape of the EM as a function of temperature. In \cref{fig-jfootem} (bottom), the H2012 and C2015 inversion results are depicted during the times of peak emission. Both distributions have similar shaped EM curves with disagreements in bins at low temperatures ($\log T_e <6.0$) and at high temperatures ($\log T_e>7.0$).

The J2 shows two distinct temperature peaks. Gaussian fitting over each of the two observed temperature peaks reveals centers at $\log T_{e}/K \sim 6.40\pm0.20$ and $\log T_{e}/K \sim 7.00\pm 0.19$. Fitting a single Gaussian covering both temperature sub ranges revealed the J2 region averaged temperature $\log T_e/K = 6.70\pm0.22$ comparable to the result of the A2013 method. Although the emission appears separated, the hot temperature component dominates and almost completely blends with the lower EM peak. 

The J3 EM curve is dominated by a peak at higher temperatures, with a Gaussian fit revealing a center at $\log T_{e}/K \sim 6.75\pm0.45$. The  temperature width is considerably higher when compared to the other two eruptions. For this more broad and high-temperature jet, the more convoluted H2012 and C2015 inversions can be approximated with A2013. 

The J6 footpoint presents a consistent double peak. Gaussian fitting over each temperature peak reveals peak EMs centered at $\log T_{e}/K \sim 6.30\pm0.15$ and $\log T_{e}/K \sim 7.00\pm 0.20$. 

We utilized the H2012 and C2015 temperature distributions to limit the EM over the above defined dominant temperature ranges. Using the same temporal windows as in the case of A2013 we calculated the geyser footpoint region averaged total EM using H2012 and C2015, finding them to be at least compatible with the A2013 determination. The dominant geyser temperatures and plasma densities recovered via the A2013, H2012 and the C2015 inversions are recorded in \cref{table-allparam}.

\subsection{Jet Outflow Emission Measures}\label{sec-aia:jet-dem}

Some differences exist between the main jet and footpoint DEM analysis. When investigating the three jet outflows, we aim to reveal spatially resolvable untwisting strands that may be heated to different temperatures. Any temporal data average would smooth out these details. Additionally, when compared to the footpoint EM, the jet body is fainter resulting in almost no filtergram intensity saturation, leading to better constrained results as hinted in Sec. \ref{sec-aia:sub-jfoot}. 

\begin{figure}[!t]
\begin{center}
\includegraphics[width=1.02\linewidth]{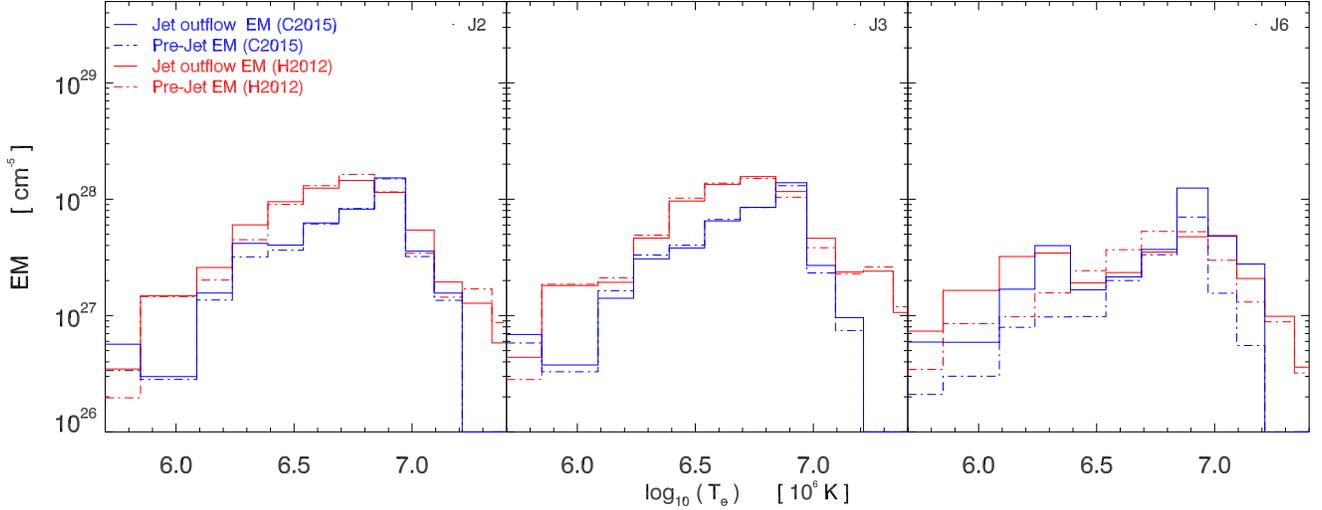}
  \end{center}
\caption{The J2, J3, and J6 EM profiles across the temperature space as derived from the H2012 and C2015 inversions. The erupting material is plotted against the background selected for each individual jet. Data was re-binned in the same temperatures intervals in order to directly compare the EM profiles.}
\label{fig-jetprofiles}
\end{figure}

We selected one time instance in which each jet outflow appears clearly separated from the footpoint. Each eruption was enclosed in a region of interest surrounding the outflowing material. The same selection was then applied to pre-jet background times.  This procedure was repeated for two additional jet frames, at $\pm12$ s before and after the original selection, revealing no substantial differences in the temperature distributions of inverted EMs. Following this setup, the detailed EM distributions are compared with pre-jet background EM distributions in \cref{fig-jetprofiles}. The A2013 results can not be directly compared.

The pre-flare background conditions are similar for all three jets. We  expect this to happen in the case of J2 and J3 which are temporally separated by about 10 min. J6, occurring 12 h later than J3, has a similar background profile. Both the C2015 and H2012 background results depict an expected quasi-uniform distribution of coronal plasma in the $\log T_e/K=[6.0,7.0]$ range with a slight prevalence of hotter plasma. This is a property of the selected region at each specific time, and such distributions are unique based on region selections that are constructed. These pre-jet profiles thus show that the geyser was tracked with sufficient accuracy during the long observing period involved. 

All three analyzed jets show particular EM distributions shapes within the selected temperature range. J2 can be characterized by three distinct temperature regions: a small but significant emission in the lower coronal range $\log T_e/K\sim [5.9,6.4]$, no significant change from background EM in the $\log T_e/K\sim [6.6,6.8]$, and a distinct EM recovered in the $\log T_e/K\sim [6.9,7.2]$ range. The J3 jet is revealed to have no significant increase in EM, or even significant decreases, in the $\log T_e/K\sim [5.9,6.7]$ region, and increased EM in the $\log T_e/K \sim [6.8,7.2]$ region. On the other hand, the J6 jet clearly presented two distinct and significant emitting structures, one lower temperature component $\log T_e/K \sim [6.1,6.5]$, and one hot emission component, $\log T_e/K \sim [6.9,7.1]$. Tn the $\log T_e/K\sim [6.4,6.8]$ range, a significant decrease in EM via H2012 is present not so much via C2015.

Are the inversion results recovered using the three methods at least compatible? J2 and J6 deviate from a single-gaussian distribution. We select J3 and interpret the A2013 inversion as an `EM weighted average temperature', representing the total of emitting material in bins that sit inside a hypothetical Gaussian function with A2013 fit parameters $a=6.28 \cdot 10^{28}$ cm$^{-5}$, $b=6.83$, and $c=0.28$. We then compare these fits with the H2012 and C2015 EM distributions. We calculate EMs of $2.81 \cdot 10^{28}$ cm$^{-5}$  and $ 2.89 \cdot 10^{28}$ cm$^{-5}$ for C2015 from H2012 respectively. Thus, at least in the case of the J3, the total A2013 EM appears overestimated by a factor of $\sim$2.

The jet outflow EMs can be compared with their footpoint counterparts. In Sec. \ref{sec-aia:sub-jfoot}, we hypothesized that the two emission peaks observed in \cref{fig-jfootem} (bottom) may correspond to a superposition of plasma resulting from the geyser footpoint and jet erupted material. Consequentially, since the jet outflow is tracked inside a region that does not contain the footpoint, the temperature distribution represented in \cref{fig-jetprofiles} should be dominated by the lower temperature plasma. This hypothesis proved erroneous as a distribution of higher and lower temperature emission could be established along all jet outflows analogous to the footpoint estimation. 

\begin{figure}[!p]
\begin{center}
\includegraphics[width=1.02\linewidth]{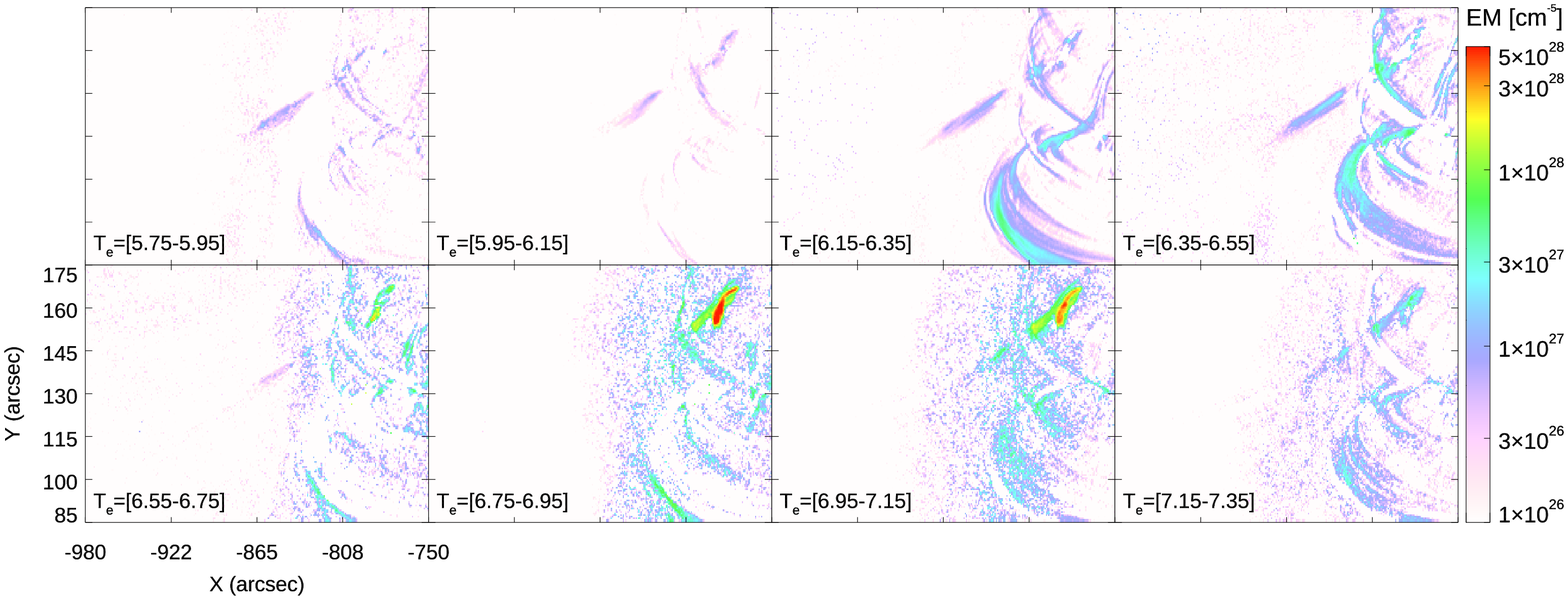}
\includegraphics[width=1.02\linewidth]{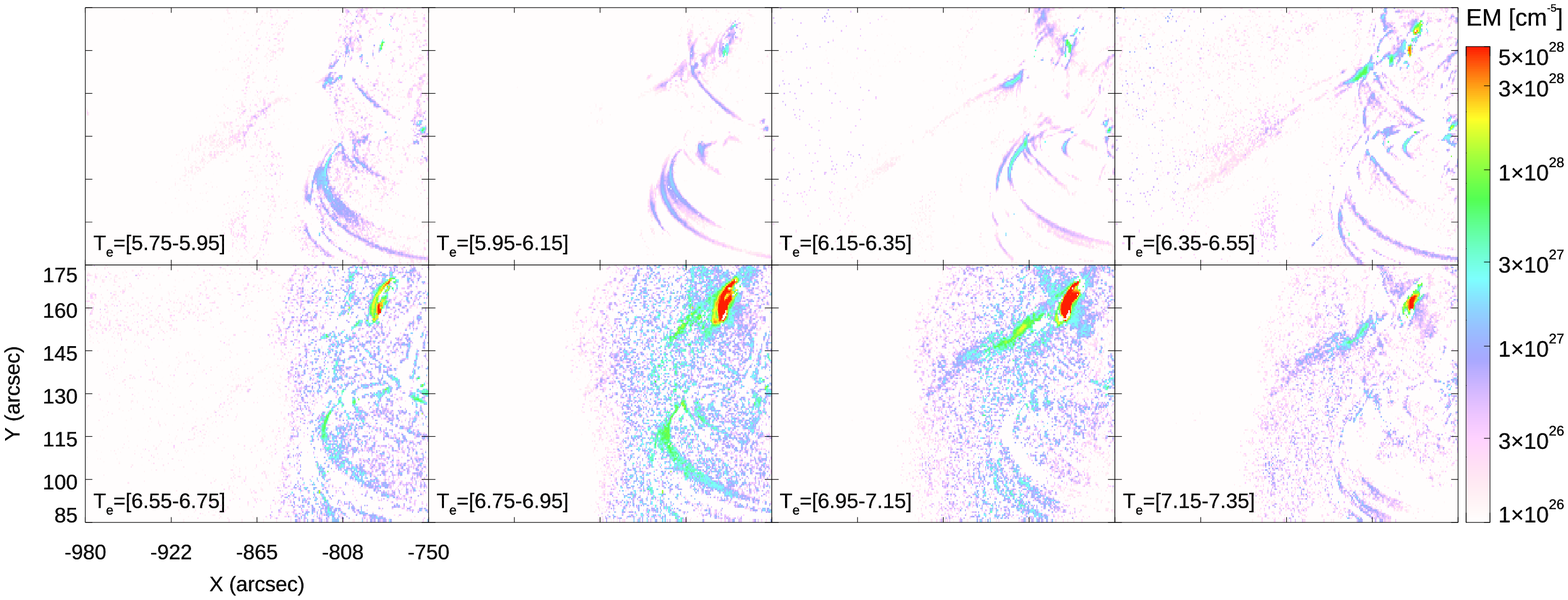} 
\includegraphics[width=1.02\linewidth]{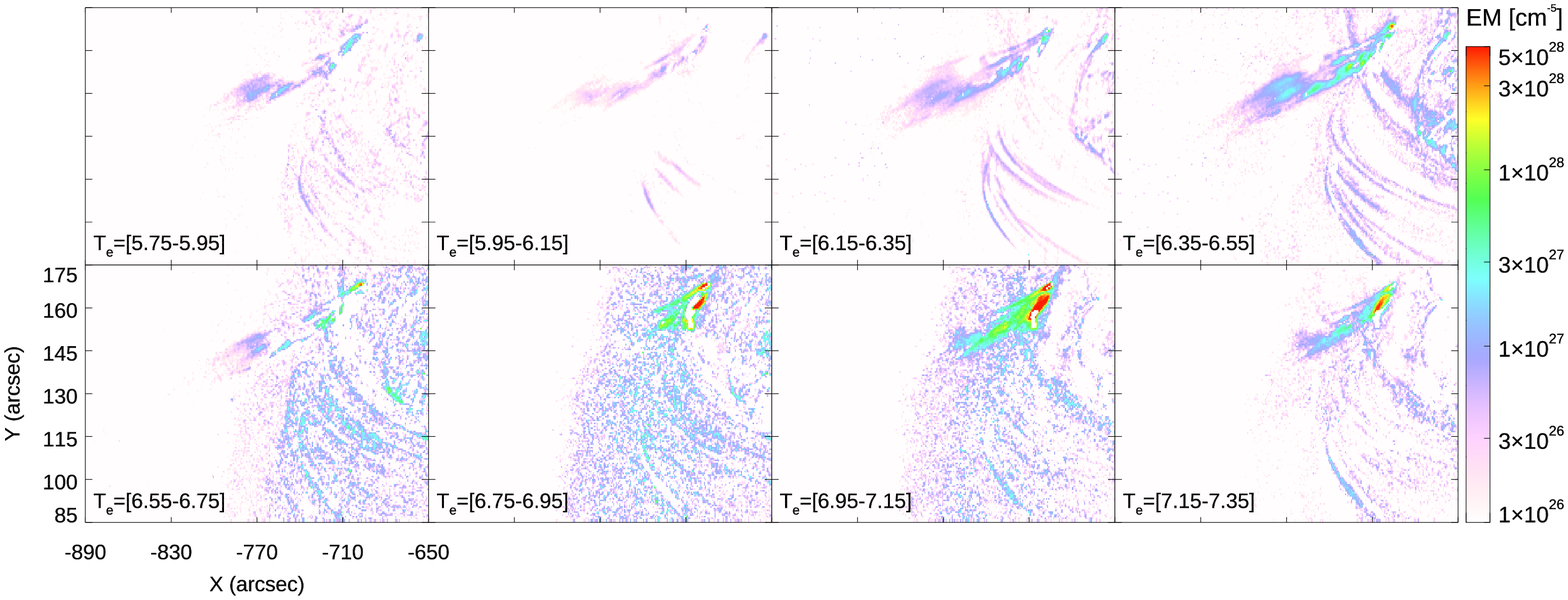} 
\end{center}
\caption{The C2015 inversion EM background subtracted maps corresponding to the erupting jet material and unique geyser footpoint presented for the J2 (top), J3 (middle), and J6 (bottom). The EM profiles are mapped across the selected $T_e$ bins revealing different strands (spires) manifesting in different temperature ranges and spatial locations in all three cases.}
\label{fig-jetemmap-cheung}
\end{figure}

The multi-thermal distributions can be alternatively explained by multiple strands, heated to different temperatures that are erupting simultaneously with overlapping emission distributions along each LOS. The multiple EM spikes presented in the geyser timeseries plots (\cref{fig-jfootem}, top) are possibly generated by successive fast reconnective events inside a blow-out type eruption mechanisms as described by \cite{moore2013} and \citet{sterling2015,sterling2016}. Further evidence is provided by the EM maps recovered via the H2012 and C2015 inversion results. \Cref{fig-jetemmap-cheung} shows the C2015 recovered EMs across all temperature bins. It is revealed that two main spatially separated strands exist for all jets; one manifesting in the low temperature intervals and one in the higher temperature range. These visible strands are morphologically unique and are separated spatially. The same conclusion can be replicated using the H2012 inversion results. We note that isolating isothermal temperatures from filter data is not straightforward \citep{judge2010}.

\begin{table}[!t]
\centering\footnotesize
\caption{The jet outflow and geyser footpoint morphological, geometrical, and inverted thermal physical parameters are presented. The morphological parameters are derived as average values when measuring discretized distances using the AIA-304\AA{}, AIA-171\AA{}, AIA-94\AA{}, and AIA-131\AA{} filters. The plasma physical and energetic parameters, $\log T_{e}$ and $n_{e}$, are derived via the A2013, H2012, and C2015 inversion approaches, taking into account the morphological parameters, geometrical approximations, and the particular assumptions and limitations of the models (see app. \ref{sec-appendix}). A filling factor $\phi=1$ is assumed.  The C2015 and H2012 distributions showed double temperature peaks for J2 and J6. Thus, the $n_{e}$ determinations correspond to each $\log T_{e}\pm \sigma_{T_e}$. } 
\label{table-allparam}
\begin{tabular}{cccccc|ccc|ccc}
  \hline\hline
&  \multirow{2}{*}{No.}&Time&Width&Height&v$_{proj}$& \multicolumn{3}{c|}{$\log T_{e}\pm \sigma_{T_e} \quad [K]$}& \multicolumn{3}{c}{$n_{e}\quad [10^{11}$ cm$^{-3}]$}  \\
  \small & &[hh:mm] &[km] &[km]&[km~s$^{-1}$]&{\color{blue}C2015} & {\color{orange}A2013} & {\color{red}H2012} & {\color{blue}C2015}& {\color{orange}A2013} & {\color{red}H2012} \\
  \hline
\multirow{8}{*}{\rotatebox[origin=c]{90}{Jet Outflow}} &   \multirow{2}{*}{   J2  }     & \multirow{2}{*}{01:14:00} &   \multirow{2}{*}{ 2574} &  \multirow{2}{*}{80064} &  \multirow{2}{*}{224} & {\color{blue} $6.00\pm0.30$}&  \multirow{2}{*}{{\color{orange} $6.58\pm 0.20$}} & {\color{red} $6.10\pm0.08$} & {\color{blue}$0.03$} & \multirow{2}{*}{{\color{orange} $0.16$}} & {\color{red} $0.05$}  \\
 &      &                                               &                                             &                                            &                                        & {\color{blue} $7.05\pm0.15$} &                                                                                        &  {\color{red} $7.10\pm0.06$} & {\color{blue}$0.07$} &                                                                         &{\color{red} $0.06$}  \\
\\
   &  \multirow{2}{*}{J3}         & \multirow{2}{*}{01:22:24}  &  \multirow{2}{*}{3706} &  \multirow{2}{*}{94323}  &  \multirow{2}{*}{192} & \multirow{2}{*}{{\color{blue} $6.90\pm0.30$}} &  \multirow{2}{*}{{\color{orange} $6.83\pm 0.28$}}  & \multirow{2}{*}{{\color{red} $6.89\pm0.29$}} & \multirow{2}{*}{{\color{blue}$0.09$}} & \multirow{2}{*}{{\color{orange} $0.13$}}     & \multirow{2}{*}{{\color{red} $0.09$}} \\       
&       &                                                &                                           &                                             &                                         &                                                                                &                                                                                           &                                                                                &                                                                       &                                                                            &                                       \\
\\
&  \multirow{2}{*}{ J6 }      & \multirow{2}{*}{13:10:24}  &  \multirow{2}{*}{4450} &  \multirow{2}{*}{66135} &  \multirow{2}{*}{295} & {\color{blue} $6.15\pm0.15$} & \multirow{2}{*}{{\color{orange} $6.54\pm 0.33$}}& {\color{red} $6.17\pm0.13$}    & {\color{blue}$0.03$} & \multirow{2}{*}{{\color{orange} $0.09$}} &{\color{red} $0.04$}  \\
&      &                                                &                                           &                                            &                                         & {\color{blue} $7.05\pm0.15$} &                                                                                      &{\color{red} $7.00\pm0.22$}     & {\color{blue}$0.05$} &                                                                         & {\color{red} $0.05$}  \\
\hline
\multirow{8}{*}{\rotatebox[origin=c]{90}{Geyser Footpoint}}&  \multirow{2}{*}{  J2  }       & \multirow{2}{*}{01:13:00}  &  \multirow{2}{*}{4978}     & \multirow{2}{*}{17889} &\multirow{2}{*}{n/a} & {\color{blue}$6.40\pm0.20$} & \multirow{2}{*}{\color{orange}$6.58\pm0.22$}  & {\color{red}$6.38\pm0.21$} & {\color{blue}$0.08$} & \multirow{2}{*}{{\color{orange}$0.18$}} & {\color{red}$0.09$}  \\
&  &                                                &   &                                            &                                     & {\color{blue}$7.00\pm0.20$}&                                                                                   & {\color{red}$6.96\pm0.18$} & {\color{blue}$0.10$}  &                                                                       &{\color{red}$0.13$}  \\
\\   
&\multirow{2}{*}{J3 }        & \multirow{2}{*}{01:19:00}  & \multirow{2}{*}{ 3926}    & \multirow{2}{*}{20972} &\multirow{2}{*}{n/a} &\multirow{2}{*}{{\color{blue}$6.75\pm0.45$}} &  \multirow{2}{*}{{\color{orange}$6.80\pm0.34$}}&  \multirow{2}{*}{{\color{red}$6.75\pm0.41$}}& \multirow{2}{*}{{\color{blue}$0.23$}} &  \multirow{2}{*}{{\color{orange}$0.32$}}&  \multirow{2}{*}{{\color{red}$0.26$}} \\    
&  &                                                &  &                                           &                                     &                                                                                &                                                                                      &                                                                               &                                                                     &                                                                       &\\
\\
& \multirow{2}{*}{J6}        & \multirow{2}{*}{13:08:24} &  \multirow{2}{*}{6150}   & \multirow{2}{*}{17220} &\multirow{2}{*}{n/a} & {\color{blue}$6.35\pm0.15$} &\multirow{2}{*}{\color{orange}$6.63\pm0.25$}  & {\color{red}$6.29\pm0.08$} & {\color{blue}$0.09$} &\multirow{2}{*}{{\color{orange}$0.28$}} & {\color{red}$0.10$}  \\
&   &                                               &    &                                           &                                     & {\color{blue}$7.00\pm0.20$} &                                                                                  & {\color{red}$6.96\pm0.12$} & {\color{blue}$0.15$} &                                                                       & {\color{red}$0.19$}  \\         
  \hline\hline
  \end{tabular}
\end{table}         

The jet eruption outflow EM and peak temperatures as recovered by the A2013, H2012, and C2015 inversions can be found in \cref{table-allparam}. The H2012 and C2015 inversions revealed double peaks in the case of J6 and J2. In the case of J2, the lower temperature component is modest. In the inverted EM maps (\cref{fig-jetemmap-cheung}), more than two strands at different temperatures can be visually observed, as hinted by the SDO-AIA filtergrams (\cref{fig-alldata}). We note that this analysis provides a simplified picture of the actual eruption configuration, where finer details are not accurately recovered in the inverted EM maps due to the spatial resolution of the filtergram observations, high solution thermal widths, low counts characteristic of integrating EM in small bins, and limitations in the ill-posed inversions. 

The above results become very important when discussing the individuality of jet eruptions. Using the basic parameters summarized in \cref{table-allparam}, we conclude that the individual eruptions, when scrutinized, are geometrically and physically unique, thus contradicting a homologous self-repeating eruption scenario. Ultimately, this analysis can not support such argument by itself. In a complementary work, we assessed the magnetic triggers responsible for this geyser, finding that at least these recurrent jets are not in fact homologous \citep{paraschiv2020}. 

\section{Analysis and Results (II): The Geyser X-Ray Energetics} \label{sec-rhessi}
\subsection{RHESSI imaging X-Ray source reconstruction}

In the standard flare picture, upwards and downwards beams of non-thermal particles are generated alongside EUV and X-Ray thermal emission. A qualitative schematic of observable small-scale flare signatures is presented in fig. 1 of \citet{paraschiv2019}. The complementary work correlates heliospheric beam propagation with multiple geysers, showing that these routinely produce upwards electron beams. Here, we further investigate the nature of the reconnective processes by means of higher energy spectroscopy, pursuing signatures of the more elusive down-streaming electron beams.
  	 
\begin{figure}[!t]
  \begin{center}
    \includegraphics[width=1.02\linewidth]{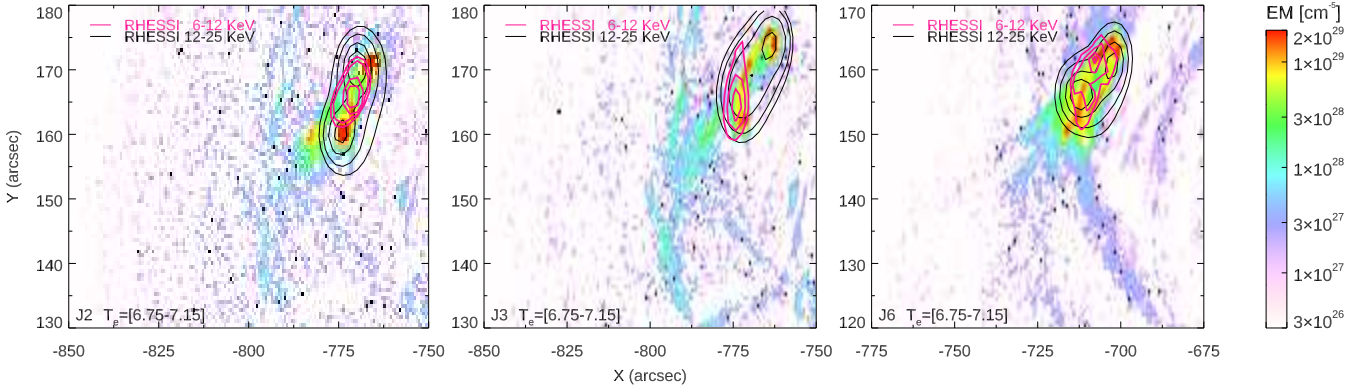}
  \end{center}
  \caption{RHESSI X-Ray emission over-plotted on the three EM maps of the jet eruptions (the J2-01:13UT, J3-01:19UT, and J6-13:08UT eruptions). Contours show the soft (magenta) 6-12 KeV and hard (black) 12-25 KeV energy channels and represent the normalized [0.3, 0.5, 0.7, 0.9] levels. The EM maps were recovered via C2015 at jet flaring times, which are temporally closest to maximum peaks in the RHESSI flux data. }
  \label{fig-rhessi}
\end{figure}

\Cref{fig-rhessi} presents the pixon X-Ray source reconstruction, as contours plotted over the EM maps. The EM maps were recovered using the C2015 inversion and the RHESSI source reconstructions used integrated signals around peak flaring times for durations of 24 s for J2, 20 s for J3, and 16 s for J6, respectively. Each EM map is summed over the higher SDO-AIA temperature range, $6-10$ MK, showed above in \cref{fig-jetemmap-cheung} to be most responsive to the jets. The three eruptions peaked in the 6-12 KeV and 12-25 KeV RHESSI energy bands, with marginal emission above background counts in the lower ($<6$ KeV) and higher energy channels ($>25$ KeV). 

From a solar plasma physical perspective, we attribute soft X-Ray emission to thermal radiation of hot loops assumed in a quasi equilibrium state and hard X-Ray emission to a thick-target bremsstrahlung process of non-thermal electrons, that are supposedly accelerated during flaring. The presence of non-thermal emission is regarded as an indicator of impulsive flaring events. In all our three particular cases distinct 12-25 KeV  vs. 6-12 KeV X-Ray source morphologies were reconstructed as seen in \cref{fig-rhessi}. It can be observed that the contoured sources do not perfectly overlap the SDO-AIA EM map footpoints. Causes for this include: (i.) Seven out of the nine RHESSI detectors were used, limiting the resolution and input data counts and thus the accuracy of the pixon determination. (ii.) The RHESSI sources were recovered with a maximum $2\arcsec$ binning. (iii.) The high longitude position of the geyser site makes it prone to geometrical and projection effects that limit the accuracy of the reconstruction. Additionally, the size of the solar disk slightly varies between the EUV and X-Ray wavelengths due to opacity differences.

The 12-25 KeV X-Ray source locations are wide and elongated, exhibiting two distinct emission locations in the case of J2 and J6, and one strong source along with a very elongated lower intensity contour, oriented towards the bottom flaring footpoint of J3. These indeed seem to qualitatively correspond to the footpoints of the flaring loops involved in the jet eruptions, as shown in \cref{fig-jfil} and \cref{fig-jetemmap-cheung}.  In the case of the  6-12 KeV emission, the sources appear smaller and seem to consistently sit between the two 12-25 KeV and EUV loop footpoints, for all three jets. This visual interpretation may lead us to attribute a hard X-Ray label to the 12-25 KeV emission and soft X-Ray association for the 6-12 KeV channel. 

Is it possible to associate the 12-25 KeV separated footpoints to impact sites of downstreaming non-thermal electron beams, that are `braked' by the lower atmosphere? The 6-12 KeV emission can be in turn interpreted as thermal emission from the reconnection heated loop top. \citet{judge2017} analyzed a presumably non-thermal small-scale flaring site \citep[`ribbon D'; ][]{testa2014}, and showed that such assumptions, although intuitive, may not always reflect local conditions, finding higher energy emission to be the result of chromospheric flare heating. Although compelling, when considering the geyser jets, the X-Ray source reconstruction is not a sufficient argument by itself.

\subsection{RHESSI X-Ray spectral analysis}

In general, X-Ray thermal emission is predominant in the low energy bands, while non-thermal emission becomes dominant at energies  $>20$ KeV. In part this is because of the $1/\epsilon^2$ dependence of the $e^--e^-$ collision time, of electrons with  energy $\epsilon$. In practice, the energy cutoff needs to be addressed on an individual basis as eruption power scaling and local conditions can skew interpretation.

\begin{figure}[!p]
  \begin{center}
       \hspace{-0.2cm}\includegraphics[width=0.329\linewidth,height=8cm]{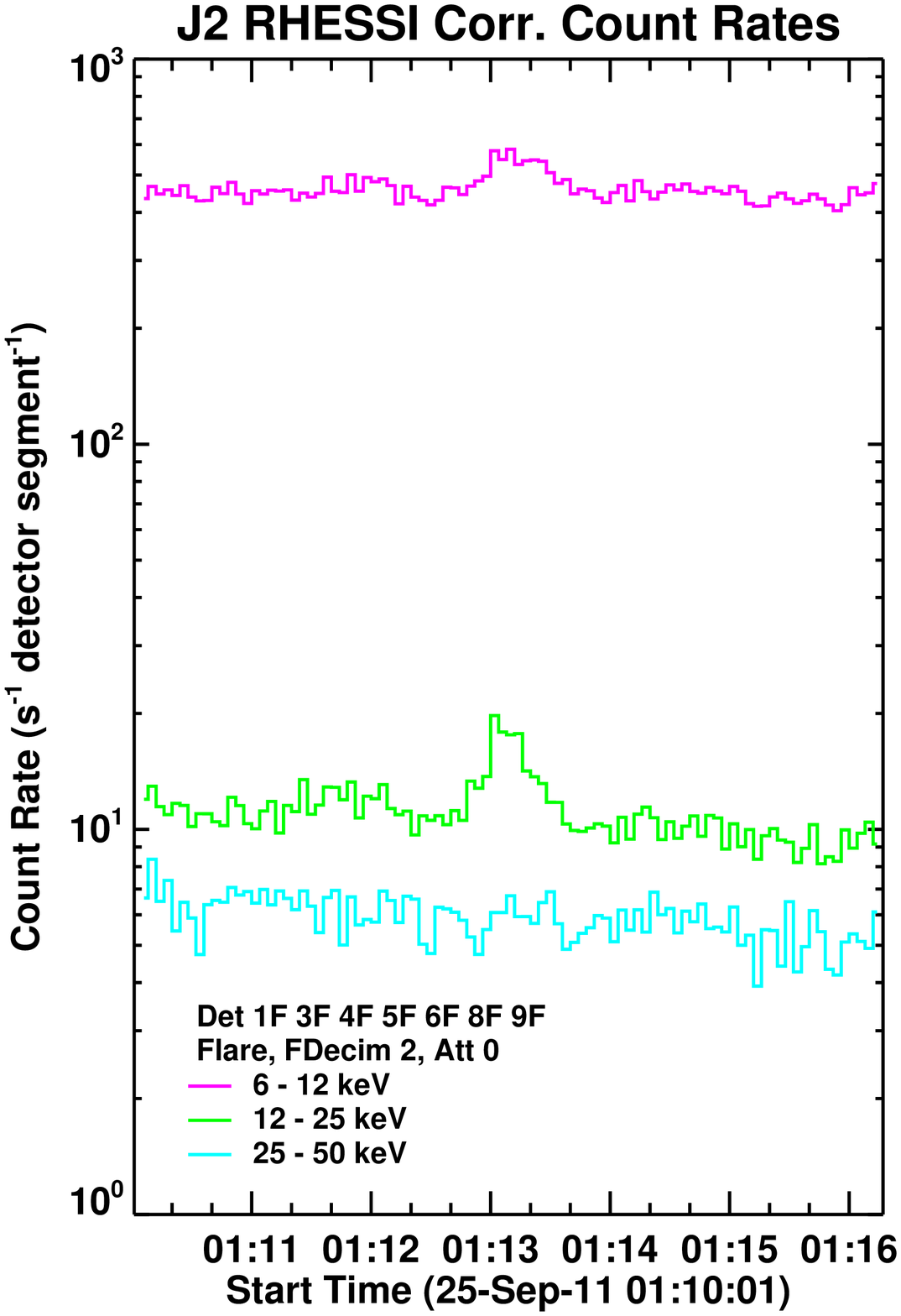}
       \includegraphics[width=0.329\linewidth,height=8cm]{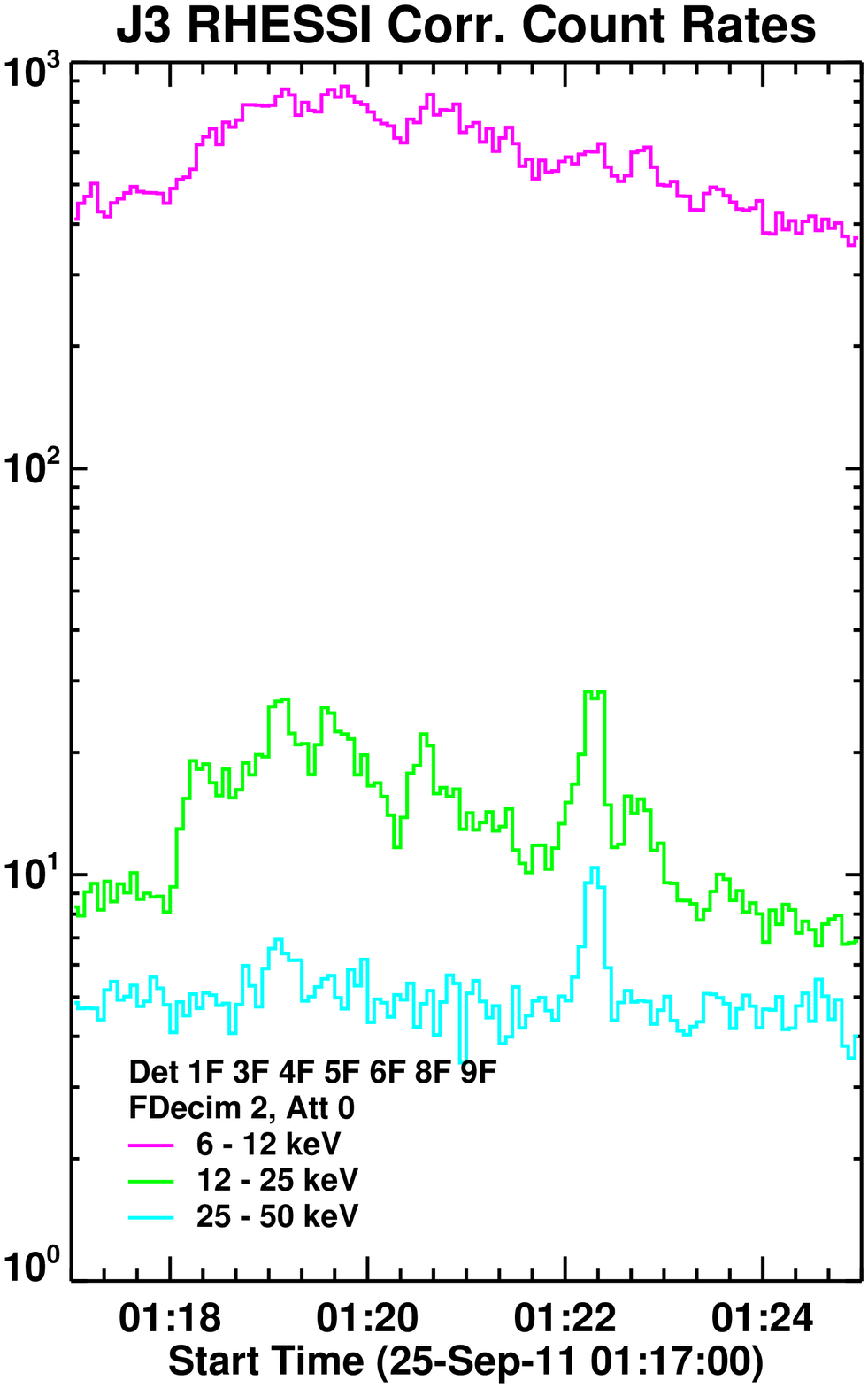}        
       \includegraphics[width=0.329\linewidth,height=8cm]{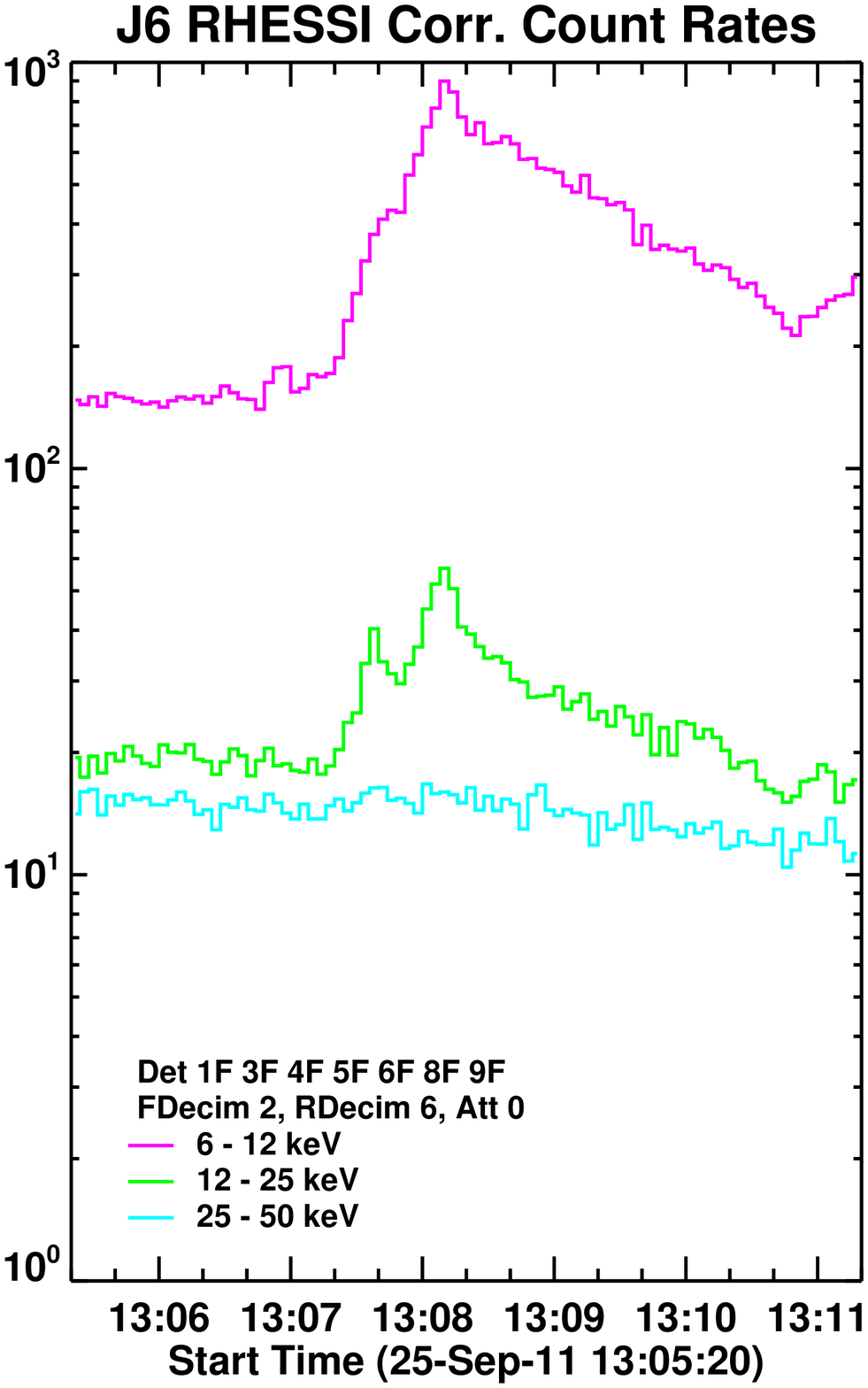}  
 
        \includegraphics[width=0.329\linewidth,height=8cm]{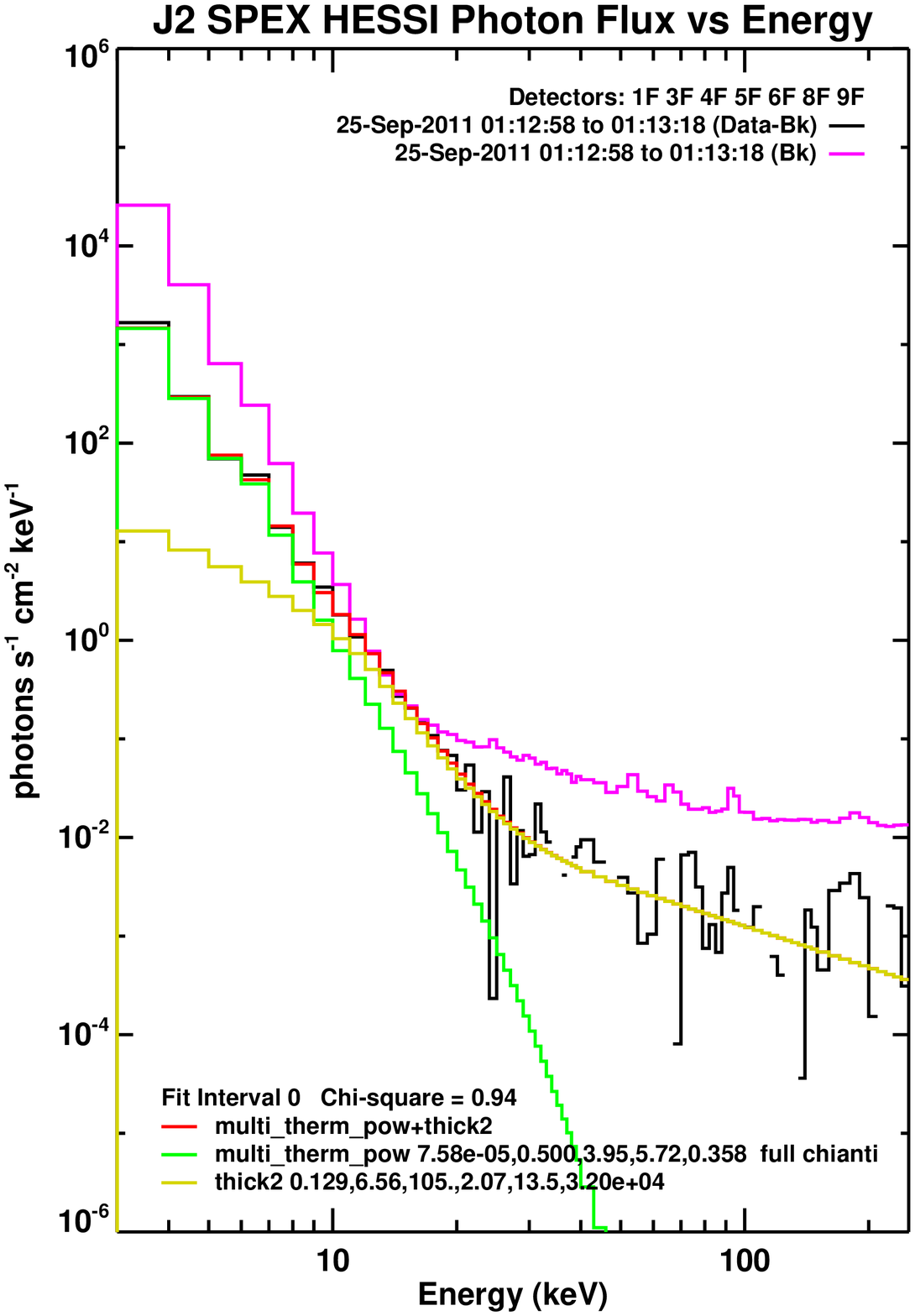}     
        \includegraphics[width=0.329\linewidth,height=8cm]{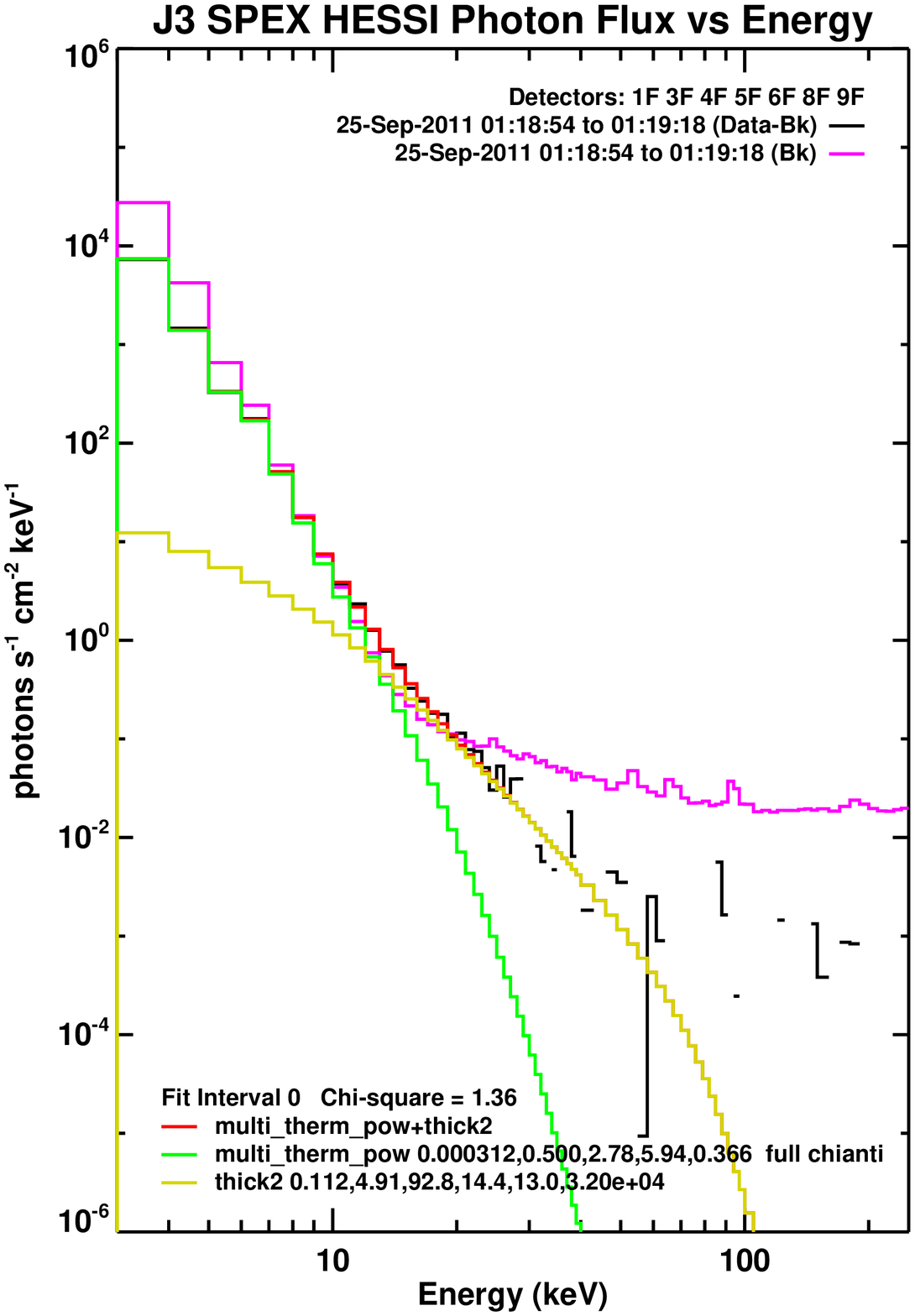}     
        \includegraphics[width=0.329\linewidth,height=8cm]{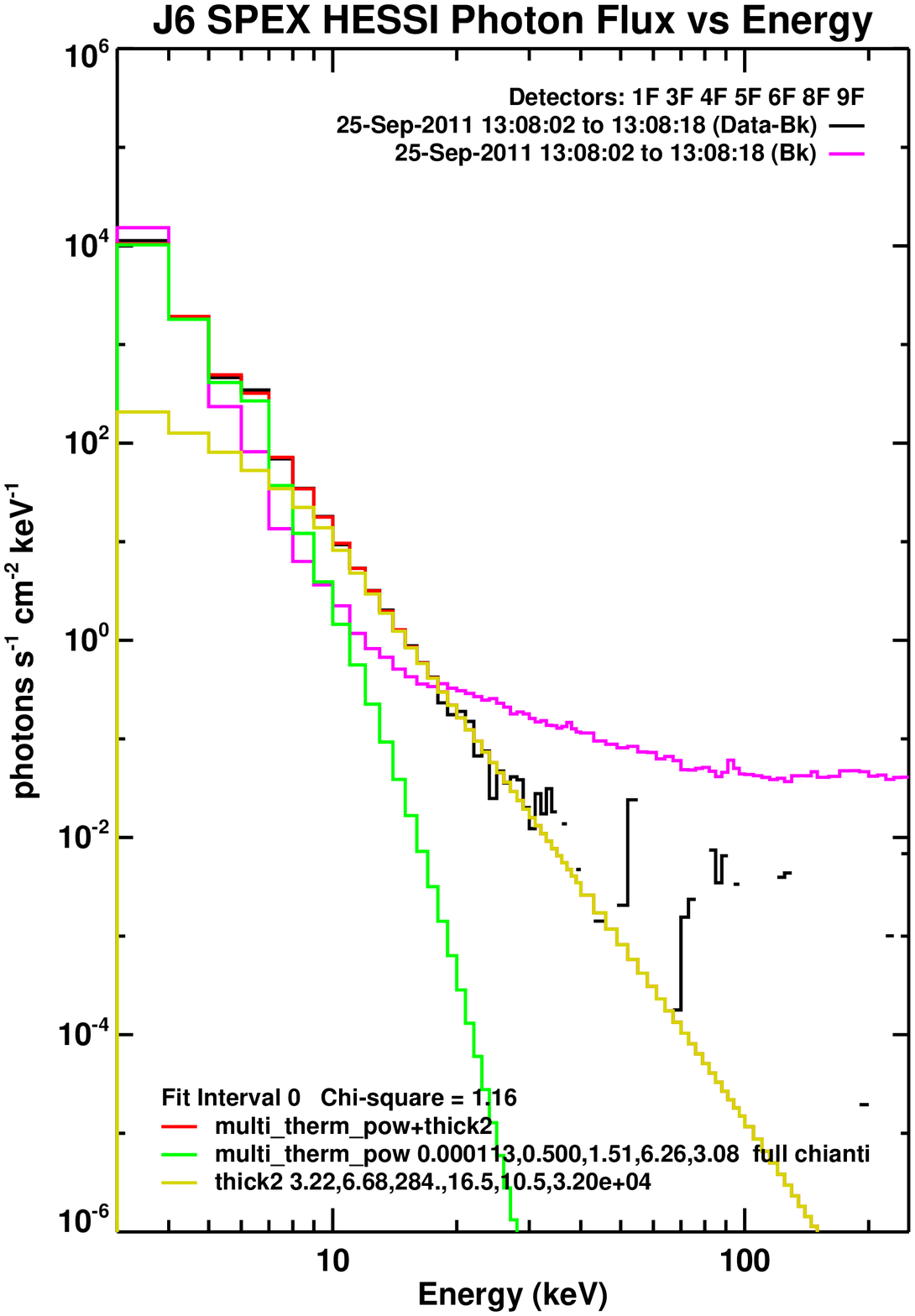}       
  \end{center}
  \caption{RHESSI X-Ray emission spectra corresponding to the J2, J3, and J6 jet eruptions. Top Panel: Timeseries plots of the total X-Ray count rates for the 6-12 KeV, 12-25 KeV, and 25-50 KeV channels. Bottom Panels: Thermal and non-thermal plasma emission models fitted on the spectral distribution in the 3-250 KeV range. No emission with respect to background could be recovered in the  25-50 KeV and higher channels, with one exception at 01:22. We focus on the results at energies $<30$ KeV. The spectra were averaged over short 16-24 s timescales corresponding to the first peaks in the photon flux timeseries of the 12-25 KeV band. }
  \label{fig-rhessi-spec}

\end{figure}

 The total RHESSI count temporal evolution is depicted in \cref{fig-rhessi-spec}, (top). The resulting lightcurves are modest when compared to larger flare events. The 25-50 KeV channel does not record counts above background levels, excepting a short 12 s peak during J3. Such a behavior is expected when resolving lower power microflares. \Cref{fig-rhessi-spec}, (bottom) shows the background subtracted photon flux spectra obtained for the three eruptions (black curve), at times correspondent to the main flaring phase. The background counts are plotted in pink. We note that the flaring time photon flux, is in general few factors higher than the background. RHESSI spectrum integration is typically done in multiples of 4s. Here we integrated the spectra for the same 24 s for J2, 20 s for J3, and 16 s for J6 temporal slots used for the source reconstructions.

The X-Ray energy spectra were fitted with a series of thermal and non-thermal plasma emission models. We aimed to best reproduce the observed spectra using one or a combination of multiple models, selecting the $<30$ KeV range where counts were significantly above the background levels.  Larger flare X-Ray emission is reliably recovered by a combination of optically thin isothermal emission and a non-thermal double power law function that models thick-target bremsstrahlung emission (\emph{vth+b\_pow}). We found that no single model was able to accurately reproduce any of the three eruption spectra.  Models based on exponential function distributions (e.g. \emph{multi\_therm\_exp}) proved less reliable in all our cases. A set of two  models was found to reliably reproduce all three eruptions. We have utilized the multi-thermal power function (\emph{multi\_therm\_pow}) (green curve) to reproduce the lower energies in combination with a double power law thick-target bremsstrahlung (\emph{fthick2}) to model (yellow curve) the higher energy spectral range. We note that this model set (red curve) yielded the best fit with  $\chi^2$ residuals of 0.94 (J2),  1.36 (J3), and 1.16 (J6). Prior to and after all jets, thermal models could fit background counts, and as expected, the non-thermal model was unable to adequately match the spectra. Qualitatively, we observe that the intersection between the two fitting model functions differ for the three jets, as analogous to the unique EM temperature distributions of each jet. 

We note that the individual fit parameters correspond to physical quantities. The counts in the higher energy range allow for an estimation of the downward beam electron flux. An X-Ray DEM measure is recovered for each jet. We compute the thermal DEM of the RHESSI sources corresponding to a peak temperature $T_e=2$ KeV, or $\log T_e/K \sim7.36$, assuming a volume approximation by utilizing, 
\begin{equation}
V=\pi\cdot  \frac{D^2}{4}\cdot S,
\label{chap:energ:sec-rhessi:eq-vol}
\end{equation}
where $D$ is the footpoint loop diameter and $S$ is the loop arc length. 

The DEMs resulting as residuals from the RHESSI \emph{multi\_therm\_pow} thermal fitting component are 0.000078 (J2), 0.000312 (J3), and 0.000113 (J6) $\cdot 10^{49}$ cm$^{-3}$ KeV$^{-1}$. These are converted using eq. \ref{chap:energ:sec-rhessi:eq-vol} to $n_e$ values of 0.03 (J2), 0.06 (J3), and 0.04 (J6) $\cdot 10^{11}$ cm$^{-3}$.

The thermal X-Ray DEMs can be compared to the EMs recovered using the EUV techniques. The findings correspond to hot plasma in the last reliable bin used for the EUV calculations. Suppose we interpret the EUV EM profiles from \cref{fig-jfootem} (bottom) to follow a similar power law decrease at higher temperatures. Assuming we allow the EUV $\log T_e/K=7.36$ results to be significant, we measure the $n_e$ of the three jets to be in the order of $0.02-0.06 \cdot 10^{11}$ cm$^{-3}$. The two independent estimations therefore appear to be at least compatible. We note that the minimum \emph{multi\_therm\_pow} fit temperatures go beyond the data range to 0.5 KeV ($\sim 6MK$.) for all three jets.   We found analytical computation of RHESSI DEMs at 0.5 KeV to not match the corresponding 6 MK EUV DEMs, where  the first are higher by factors 7-10. We concluded that at least in our case, the fits are not reliable outside of the RHESSI measurement range.

In a larger context, we note that although the geyser, and more generally the footpoints of jets, are of modest size and energy, they can be isolated in a full disk X-Ray signal integration. It is worth investigating how many automatically detected individual microflares studied by \citet{hannah2008} have initiated jets? We discuss in Sec. \ref{sec-concl:sec-cor} the implications arising if numerous jets inject mass to the slow solar wind flux! 

\section{Discussion}\label{sec-concl}

\subsection{The differential emission measures of recurrent active region jets}\label{sec-concl:sec-dem}

 The jet eruptions have been described in terms of the EM inferred from the SDO-AIA observations. The recovered parameters from the three inversion schemes are in sufficient agreement  given the described assumptions, methods, and observational constraints. We stress that plasma inversion methods are fundamentally limited, representing mathematical models that contain significant subjectivity when addressing the ill-posed problem \citep[see discussions in ][]{craig1986,judge1997,aschwanden2015}, and \citep{cheung2015}. Even in the best particular conditions, EM inversions should be considered just approximations of plasma physical conditions and need to be corroborated with independent and complementary observations and modeling. 

\citet{testa2011} found that, in the case of microflares and nanoflares, the DEM is characterized by multi-thermal plasma; an expected $\log T_e/K=6.30$ component and a second significant  hot plasma $\log T_e/K \sim 7.00$ contribution. The authors argue that this property is compatible with existing nanoflare models. We have herein observed this characteristic for both geyser footpoint (\cref{fig-jfootem}) and jet eruption (\cref{fig-jetprofiles}). We found two distinct peaks, one around $\log T_e/K=6.2-6.4$, and one stronger component around $\log T_e/K \;6.8-7$. 

The lower temperature range estimates are consistent with the results of \citet{mulay2016}, who studied $20$ jets from multiple sites that span across multiple years of solar activity. The authors did not address higher temperature emission where  $\log T_e/K>7$ due to concerns that existed at the time with the accuracy of the solution. Based on the consistent results from our coupled SDO-AIA and RHESSI observations, we argue that the recent improvements in the CHIANTI database allow to constrain higher temperatures SDO-AIA observations, at least in this case. The high temperature emission peaks are also compatible with the \citet{moreno2008} and \citet{moreno2013} MHD results which predict hot $\log T_e/K \sim 7$ emission.  We note that the $\log T_e/K>7.3$ data is still highly unreliable. When discussing recovered EM and correspondent $n_e$ determinations our typical $n_e=3-9\cdot 10^9$ cm$^{-3}$ is comparable to the \citet{mulay2016} determinations of $n_e=2-11\cdot 10^9$ cm$^{-3}$ where we have compared only our low temperature component. On the simulation side, \citet{moreno2013} reported plasma density estimates of $n_e\sim 10^9$ cm$^{-3}$, noting that in this case, the authors were modeling a blowout coronal hole jet. We note that the density of erupting material should be a unique property of each jet or geyser.

Both this work and \citet{mulay2016} estimations arbitrarily selected a `safe' overestimated filling factor $\phi$ = 1, as EUV observations can not be solely used for accurate estimations of $\phi$. In a subsequent comprehensive study, \citet{mulay2017} brought together SDO-AIA, Hinode XRT, and Hinode EIS observations to analyze one coronal jet. One important result is the more extreme $\phi$ = 0.005, obtained via the density sensitive \ion{Fe}{12} ratio forming at $\log T_e/K = 6.30$. \citet{chifor2008b} reported similar results. \citet{mulay2017b} used IRIS observations to calculate a $\phi$ = 0.1 in chromospheric regions. \citet{judge2000} addressed spectroscopic filling factors of the transition region for both homogeneous and and non-homogeneous plasma conditions finding $\phi$ = 0.12-1 to match observations. In jet MHD simulations \citet{moreno2013} found $\phi$ = 0.2. The filling factor conundrum remains an active issue and source of significant uncertainty.  

Could there be more than two erupting components associated to jet eruptions?
\citet{mulay2017} proved the existence of multi-thermal plasma components in a jet eruption using a technique involving emission line isolation from the AIA filtergrams finding consistent \ion{Fe}{18} emission. A noteworthy problem that may arise when interpreting DEM observations is the uncertainty in the inversion of the AIA-94\AA{} filter that has complex multi-temperature plasma components \citep[][]{delzanna2013}. A solution may consist in isolating the \ion{Fe}{18} emission in the AIA-94\AA{} channel as proposed by \citet{warren2012}. \citet{mulay2017b} used lower height IRIS observations in order to further constrain a lower temperature emission component. Although they recovered hot flaring ions, \citet{mulay2017b} concluded that hot $>8$ MK emission is unlikely for jet eruptions, at least in their case. However, microflare sites have been shown to exhibit $>8$ MK thermal emission \citep{hannah2008}. Our geyser exhibits similar hot RHESSI thermal emission. We note that the high variability of jet properties may allow both interpretations to coexist. If in the case of the jets studied by \citet{mulay2016,mulay2017b} there is no discernible hot emission, one can interpret that there is no sign of impulsive reconnection occurring. Another class of jets can maybe be found by exploring this facet. 

Hot flaring $>8$ MK loop emission has been extensively observed and modeled \citep[see review; ][]{reale2014} at both large and small scales. All three inversions used herein were validated for hot flaring emission by their literature sources.  As stressed herein, DEM techniques are partly decoupled from the studied physical system and suffer from limitations. The Hinode-XRT inversion method, $xrt\_dem\_iterative2$ that is transformed and used for AIA observations by \citet{mulay2017b} has been shown to underestimate DEMs in the case of synthetic data \citep{hannah2012,aschwanden2015}, possibly corroborating the discovery of only lower temperature emission. We have shown herein, that our events, and recurrent jet inducing sites in general, exhibit high temperatures with a substantial EM increase at high temperatures.

Interpreting our observations requires a multi-thermal hypothesis involving multiple strands, heated to different temperatures that are erupting almost simultaneously. From the SDO-AIA filtergram timeseries, we can distinguish multiple strands, that are erupting simultaneously. The C2015 inversion EM maps (\cref{fig-jetemmap-cheung}) of the three jets reveal at least two main morphologically different strands that are spatially separated. Radially, they appear at slightly different heights at a single timestep. We draw attention to the \cref{fig-jitot} fluxes, where for all three analyzed eruptions, multiple short successive flaring peaks are seen. Multiple smoothed out peaks are also observable in the EM timeseries of the geyser footpoint (\cref{fig-jfootem}, top). The erupting strands can thus be interpreted as multiple short succession flaring events. We can discuss these in terms of current jet eruption models. For example, blowout minifilament eruptions involving subsequent reconnection events that seemingly give rise to jets and heliospheric manifestations such as switchbacks, were hypothesized in a series of papers   \citep{sterling2015,sterling2016,sterling2017,panesar2016,neugebauer2021}. The authors proposed magnetic cancellation across a neutral line to be the fundamental process that drives jets. Our detection of subsequent flaring events indirectly support the minifilament eruption hypothesis. 

 One interpretation of quasi periodic pulsations (QPP) observed in large scale flares and even stellar flares can explain our observations.  \citet{hayes2016} studied the bursty nature of the reconnection from multi-wavelength QPPs occurring during the impulsive phase of X class flares. The AIA signal timeseries in fig. 1 by \citet{hayes2016} qualitatively corresponds to our \cref{fig-jitot} if we disregard the substantial difference in power and time scaling. The authors interpreted the observed X flare QPP  as episodic particle acceleration and plasma heating in the reconnecting flux tubes.  This is further supported by the bursty hard X-Ray signal in \cref{fig-rhessi-spec} and the radio data presented in \citet{paraschiv2019}. On the other hand, such an association is questionable as QPP events are not fully understood, and multiple alternative interpretations have been offered \citep[see review; ][]{nakariakov2009}. \citet{nakariakov2018} interpreted QPPs in radio data of a microflare site as a superposition of multiple harmonics of oscillations, acknowledging that the interpretation is not unique. In our case, further insight is hindered by the weak X-Ray emission signal (Sec. \ref{sec-rhessi}) and the short lifetimes of microflares.  
 
\subsection{Jet energetics in a coronal context}\label{sec-concl:sec-cor}

 The coronal implications of jet eruptions have been debated extensively. The contribution that jets may have to the slow solar wind flux, or the influence in coronal heating remain open questions in the community \citep[e.g.][]{raouafi2016}. The DEM analysis in Sec. \ref{sec-aia} was a necessary step in order to evaluate the energetic output of AR jets.  The jet energy budget can be estimated as a sum of separate energy fluxes,
  \begin{equation}
F = F_{kin}+ F_{pot} + F_{th}.
  \label{chap:energ:sec-concl:eq-flux}
  \end{equation} 
where the three components represent the kinetic, potential, and internal energy flux estimations \citep{pucci2013, paraschiv2015}. We did not include additional terms like a radiative loss flux or an Alfv\'{e}nic wave flux to this work. The radiative loss flux can be computed from EUV EM maps \citep[e.g.][]{aschwanden2005,gilbert2013,schad2021} but was shown in different circumstances to be systematically more than one order of magnitude lower than kinetic fluxes \citep{pucci2013,gilbert2013,paraschiv2015} and thus considered negligible. Line spectroscopy is needed to accurately account for the Alfv\'{e}nic wave flux \citep[e.g.][]{kim2007}. Computations of thermal conduction timescales are hard to constrain due to cadence when using SDO-AIA observations only. Therefore F is an underestimate of energy release in AR jets. 

The three flux quantities can be approximated by:
   \begin{equation}
F_{kin} =\frac{1}{2}\cdot n_e\cdot m_H\cdot v^3 \qquad\quad\quad [erg\cdot \text{ cm}^{-2}\cdot s^{-1}],
  \label{chap:energ:sec-concl:eq-flux-kin}
  \end{equation} 
     \begin{equation}
F_{pot} =n_e\cdot m_H \cdot g \cdot H \cdot v \qquad\quad [erg\cdot \text{ cm}^{-2}\cdot s^{-1}],
  \label{chap:energ:sec-concl:eq-flux-pot}
  \end{equation}
     \begin{equation}
F_{th} =\frac{\gamma }{\gamma-1} \cdot n_e\cdot k_B\cdot T_e \cdot v \qquad [erg\cdot \text{ cm}^{-2}\cdot s^{-1}].
  \label{chap:energ:sec-concl:eq-flux-th}
  \end{equation}
  
Here, $n_e$ represents the averaged outflow density, $v$ is the outflow speed of the erupting plasma. $H$ is the height of the jet, $g$ represents the gravitational acceleration of the sun ($g=274.13$ m s$^2$ ) at $1R_{\odot}$, and $\gamma=\frac{5}{3} $ represents the ratio of the specific heats, assuming a monoatomic gas. The energy flux output of the three jets is presented in table \ref{table-mainenerg}. To compute the flux components we have used the physical parameters resulting from all EUV inversions. In the case of multi-thermal contributions we have summed the two components. The results are then compared with the polar jet estimates of \citet{pucci2013} and \citet{paraschiv2015}. For comparison, the coronal flux losses are $\sim$10$^7$ $\text{erg}\cdot \text{cm}^{-2}\cdot \text{s}^{-1}$ \citep{withbroe1977}.

\begin{table}[!h]
\centering\footnotesize
\caption{Comparison of energetic flux components in units of $10^8\;erg\cdot \text{cm}^{-2}\cdot s^{-1}$.\\ The total fluxes are sums of the averages between the A2013, C2015, and H2012 derived components listed in Table \ref{table-allparam}. The \citet{paraschiv2015} estimation represents an average over 18 events. } 
\label{table-mainenerg}
\begin{tabular}{ccccccccccc}
\hline\hline
Event or Source &\multicolumn{3}{c}{$F_{kin}$} & \multicolumn{3}{c}{$F_{pot}$} &\multicolumn{3}{c}{$F_{th}$} & $F$\\
			          & A2013 & C2015 & H2012 & A2013 & C2015 & H2012 & A2013 & C2015 & H2012 \\
\hline
J2                    & 1.50 & 0.94 & 1.03 & 1.31 & 0.82 & 0.90 & 4.70 & 6.30 & 6.33 & 7.9\\ 
J3                    & 0.77 & 0.53 & 0.53 & 1.07 & 0.75 & 0.75 & 5.82 & 4.74 & 4.63 & 6.5\\
J6                    & 1.93 & 1.72 & 1.93 & 0.80 & 0.71 & 0.80 & 3.18 & 6.14 & 5.69 & 7.6\\ 
\hline
\citet{paraschiv2015} & \multicolumn{3}{c}{$0.01$}  & \multicolumn{3}{c}{$0.01$} & \multicolumn{3}{c}{$0.07$}  & $ 0.1$\\
\citet{pucci2013}     &  \multicolumn{3}{c}{$0.30$}  & \multicolumn{3}{c}{$0.03$} & \multicolumn{3}{c}{$0.17$}  & $0.5$\\
  \hline\hline
  \end{tabular}
\end{table}       

The energetic flux components are found to vary across the different jets. Across all five examples, the $F_{pot}$ term appears in general to be weaker. With the noted exception of the \citet{pucci2013} jet, the $F_{th}$ term appears to be 2-5 larger than $F_{kin}$, dominating the partition. We find that the jet EM profiles depicted in \cref{fig-jetprofiles}, where a substantial quantity of hot coronal plasma is being ejected, to be in agreement with these flux partition estimations. 

The J2, J3, and J6 parameter estimates are obtained using the EUV DEM, while the \citet{pucci2013} and \citet{paraschiv2015} results are obtained using Hinode XRT inversions. This is relevant in the context of a comparison. Multiple works \citep{su2018,schmelz2015,wright2017} claim a calibration issue leads to a difference of factor $\sim2$ between SDO-AIA and Hinode XRT instruments and proposed a scaling of the X-Ray data. Concurrently, works as C2015 (synthetic data) and \citet{hanneman2014} and \citet{mulay2017} (observational measurements) show that combining XRT with AIA observations generally improve inversion solutions. The CHIANTI database has also significantly improved in recent years. Such problems are not necessarily related to a calibration issue as DEM estimations are in general more subjected to limitations in the inversion scheme or observations used. In our case, we show that $\log T_e/K\sim7.3$ EM retrieved via C2015 returns an almost identical density estimates as the thermal X-Ray DEM fit.

From a flux perspective, the AR11302 geyser jets are more than one order of magnitude stronger than polar coronal hole jets. As shown in \citet{paraschiv2019}, the AR11302 geyser is of medium size when compared to other geysers. The polar jet contribution to coronal hole heating was shown by \citet{paraschiv2015} to be significant, but insufficient by more than an order of magnitude to explain the total coronal heating rate. Despite the higher net flux estimates for AR jets, a corresponding estimation is not valid as the formation region is topologically different.

\citet{torok2016} and \citet{cranmer2017} debate the modest heliospheric influence of polar coronal hole jets. Could the difference in scale between polar and AR jets fill the missing energy and mass release? \citet{shimojo1996} and \citet{shimojo2000} show that most Yohkoh-SXT jets occur near or in active regions. On the other hand, statistics of a very large number of events \citep{paraschiv2010} recorded at heliospheric heights($>1.5R_{\odot}$) by the twin STEREO coronagraphs showed that a overwhelming proportion of white-light jets were associated with the two polar coronal holes.  The \citet{paraschiv2010} study is centered around the solar minimum between the 23 and 24 cycles, while the \citet{shimojo2000} study is performed on an ascending phase of activity. Similarly, the \citet{paraschiv2015} XRT jets were recorded in polar coronal holes during the extended minimum period. The main parameters calculated in the \citet{paraschiv2015} were cross-checked using MHD simulations \citep{torok2016,lionello2016} and taken into account in models of mass and energy injection to the solar wind outflow. 

Thus, any determination of AR jet flux and mass outflow needs to be addressed in the same context of the global solar activity. We argue that AR jets are a relatively scarce phenomena when compared to polar jets, and probably can only offer momentary inputs to the solar wind in the form of transients. The solar wind stream is currently discontinuous e.g. the sources are nor fully resolved, in regions close to the solar surface. \citet{neugebauer2012} showed that microstreams in the solar wind were property-wise correlated to polar coronal hole jets. It is much more challenging to prove such a connection for AR jets and geysers as the heliospheric connectivity should not be taken for granted as in the case of polar jets. Our geyser dataset viewed in the context of the heliospheric travel of electron beams along `open' fluxtubes (e.g. fig. 1, \citealt{paraschiv2019}) provides information for constraining solar wind parameters. \citet{parenti2021} use combinations of in-situ and remote data to show that jets can be indeed tracked through the heliosphere. A tracking of such geyser ejecta up to in-situ particle flux detectors as PSP and Solar Orbiter may prove extremely fruitful. 

In conclusion, we hypothesize that although very energetic, AR jets lack the ubiquitousness that polar jets exhibit, limiting their potential influence on heliospheric energetics and dynamics. 

\subsection{Microflares and downwards acceleration of particles }\label{sec-concl:sec-mf}

The main coronal drivers can manifest in very wide energy range (nanoflare-microflare-flare), with non-linear power scaling. Can our geyser be considered a typical microflare site? The jets are found to be overwhelmingly stronger when compared to the polar counterparts that are also attributed to microflare reconnection. When compared to typical microflares where jets are not always detected, the geyser flaring episodes appear to be more impulsive. An analogy to standard flares may exist. As flaring events can be eruptive or confined based on local conditions, such jets can be compared to ribbon heating generated by lower atmosphere microflares and nanoflares. The nanoflare scale is usually reserved for more modest events \citep{judge1998,testa2014,tian2014,bharti2017,tian2018}.

From a thermal perspective, X-Ray spectral analysis is performed during peak X-Ray emission and EUV flaring (\cref{fig-rhessi-spec}). The EUV DEM profiles recovered in \cref{fig-jfootem} (bottom) appear to sharply decrease towards the high temperature regions. The sharp decrease is reported in microflare studies \citep{inglis2014,kirichenko2017}, where \citet{inglis2014} deduce that a Gaussian DEM can not jointly fit a combined SDO-AIA and RHESSI DEMs and propose a simple uniform DEM that has a high cutoff temperature. Although this assumption seems to solve the punctual issue of `fusing' the data, important information on the microflare source might be lost via the intrinsic smoothing in certain situations. For this dataset, a power law thermal model provided the best fit of the RHESSI spectra.

\citet{hannah2008} offer a comprehensive statistical study of automatically detected RHESSI microflares, processing over 25000 events. The authors found X-Ray EMs on the order of $10^{45}-10^{47}$ cm$^{-3}$, in volumes of about  $10^{25}-10^{27}$ cm$^{3}$ with dissipated thermal energies $E_{th}$ of $10^{26}-10^{30}$ erg. \citet{wright2017} used NuSTAR \citep{harrison2013} observations to find thermal energies $E_{th}$ = $9\; 10^{27}$ erg for an impulsive microflare.

To calculate $E_{th}$ for the RHESSI thermal emission, we have assumed the filling factor $\phi=1$ (see eq. \ref{sec-appendix:eq-neapprox}) and utilized
      \begin{equation}
E_{th}=\frac{\gamma}{\gamma-1}\cdot n_e\cdot k_B\cdot T_e\cdot V,
  \label{sec-concl:eq-therm}
  \end{equation} 
where the coronal plasma was approximated to a monoatomic gas, with $\gamma$ = 1.66. $V$ represents the emitting volume, and $n_e$ is the volume reconstructed local plasma density from the RHESSI DEMs.  We note that our employed specific heat factor of $\gamma /( \gamma-1) \sim$2.5 is slightly different from the 3 factor used generally in the literature \citep{dejager1986,hannah2008,wright2017}.

We recovered RHESSI geyser EMs in the of $10^{45}$ cm$^{-3}$ in a volumes of $\sim 10^{25}-10^{26}$ cm$^{3}$ and computed X-Ray thermal energies $E_{th}$ = 2.04 10$^{27}$ erg (J2), 4.42 10$^{27}$ erg (J3), and 2.41 10$^{27}$ erg (J6). Thus, these high-energy thermal energies are found to be compatible with the results of \citet{wright2017}, and  match the lower limits of the thermal microflare power described by \citet{hannah2008}. EMs and footpoint thermal energetics of both the AR11302 geyser and the \citet{hannah2008} dataset need to be considered as upper limits due to the $\phi=1$ assumption.

The standard flare picture also envisions downward particle acceleration, where non-thermal electron beams stream towards the newly reconnected flare footpoints resulting in chromospheric evaporation which in turn thermalizes. X-Ray emission source morphologies of the three jets were reconstructed (\cref{fig-rhessi}). Two distinct morphologies are found in hard and soft X-Ray energy bands. The distinct source morphology along with the X-Ray footpoint separation is a first indicator that we are indeed observing particle acceleration.

\begin{table}[!b]
\centering\footnotesize
\caption{Comparison of non-thermal emission estimates. Not all parameters were computed in each source.} 
\label{table-nontherm}
\begin{tabular}{ccccc}
\hline\hline
\multirow{2}{*}{Event or Source} &$\delta$ &$E_c$ & $F_e$ &$P_N$                       \\
                           &   & [KeV] & [$10^{35}$ $e$ s$^{-1}$] & $[\text{ erg s}^{-1}]$ \\
   \hline
J2                       &  6.56  &  13.5  & 0.13  & $3.42\,10^{26}$\\
J3                       &  4.91  &  13.0  & 0.11  & $3.07\,10^{26}$\\
J6                       &  6.68  &  10.5  & 3.22  & $6.57\,10^{27}$\\
\citet{hannah2008}       &  4-10  &  9-16  &  --   & $10^{25}-10^{28} $\\
\citet{inglis2014}       &  --    &  9-14  &  --   & $10^{25}-10^{26} $\\  
\citet{wright2017}       &  $>$7  &   7    &  --   & $10^{25}-10^{26} $\\  
\text{\citet{testa2014}} &  --    &  10    &  --   & $10^{24} $\\    
  \hline\hline
  \end{tabular}
\end{table}

The power $P_N$ of the non-thermal electrons that manifest at energies higher than the non-thermal cutoff $E_c$ was determined using eq. \ref{sec-concl:eq-nontherm} (see \citet{brown1971,hannah2008,wright2017}). The $F_e$ term represents the total non-thermal electron flux, and $\delta$ represents the power law spectral index. We have extracted the function parameters from fitting the $fthick2$ thick-target model (see \cref{fig-rhessi-spec}) and included them in \cref{table-nontherm} along with literature estimates. The results of \citet{testa2014} from model driven constraints on non-thermal beam scenarios applicable to nanoflares are also presented for scale comparison. Not all parameters are derived in all studies as each used different assumptions approximations. Particularly, the non-thermal cutoff $E_c$ is shown by \citet{hannah2008} to be difficult to estimate, as the expected range is affected by thermal emission. The authors chose an analytical alternative, which we also adopt:
      \begin{equation}
P_N(>E_c)=1.6\cdot 10^{-9}\cdot E_c \cdot F_e \cdot \frac{\delta-1}{\delta-2}.
  \label{sec-concl:eq-nontherm}
  \end{equation}

We find the geyser to match a non-thermally emitting microflare picture. Issues with RHESSI sensitivity might be significant. \citet{hannah2008} document issues in the quantitative estimation of non-thermal properties of less intense microflares. Only 15\% of their events were associated with identifiable non-thermal emission. Of importance is the fact that the lack of quantitative non-thermal emission fitting was probably not due to a lack of particle acceleration, but more probably due to the high uncertainties in fitting non-thermal components. In our case, although we consider the non-thermal fits as trustworthy, the fit residuals are generally $-2<\chi^2<2$ in the individual bins above $E_c$ energies in all three jets. The power in non-thermal electrons $P_N$, are considered lower limits as the $E_c$ cutoffs are considered upper limits.

In order to evaluate the chromospheric evaporation hypothesis a comparison of the power ratio between the resulting thermal heating and presumably prior energy injection via thick-target bremsstrahlung is required. We qualitatively found that thermal energies are substantially higher than their non-thermal counterparts, as analogous to \citet{hannah2008} and \citet{inglis2014}. Based on the fact that our thick-target model fitting coefficients are uncertain and have a dependence on the low observed photon flux, we could not resolve such fine details for this dataset. Thus, our data can not pinpoint chromospheric evaporation as the main process that drives electron thermalization as envisioned by the standard flare model. Similar conclusions are found by \citet{inglis2014} who offer alternative scenarios that may explain their microflare dataset, where their events appear less impulsive than the geyser analyzed here.

The geyser observations at peak flaring time are fitted by both thermal and non-thermal X-Ray emission models, showing evidence of downwards particle acceleration in the case of jet-inducing microflares. Jet reconnection is thus brought closer to the standard flare model. RHESSI is one of the most successful solar missions up to date, massively helping us advance our knowledge of flare energetics and particle acceleration for the last two decades. A new mission focused on high energy spectroscopy is highly needed to help settle the still ongoing issues of small-scale flaring.

\section{Summary} 

In conclusion, we identified a peculiar penumbral site in AR11302 that underwent multiple magnetic reconnection events and was the main trigger of recurrent solar jets. We entitled this site a  “Coronal Geyser”. We compared and cross-validated multiple inversion and reconstruction techniques for EUV and X-Ray plasma and then estimated the physical properties (e.g. temperature, density, energy flux contributions, non-thermal power, etc.) of the plasma simultaneously at the base geyser and along the jets outflow. Evidence is presented in support of cool and hot thermal emission via EUV DEMs, along with thermal and non-thermal emission of jets via source reconstruction and spectrographic analysis of X-Ray data. The main summary points are:
 
$\bullet$ The averaged background and footpoint total EMs along the full EUV corona temperature range for all eruptions is derived. The three discussed EM inversion methods give similar results, but are not in total agreement. We note that the C2015 EM results returned lower counts and that the A2013 method is not suited for multi-temperature EM distributions that are not well represented by a gaussian fit. Since the filling factor can not be directly estimated using just SDO-AIA observations, we have chosen a unitary factor. We thus acknowledge that our EM derived parameters are most probably an overestimation.

$\bullet$ When observed via individual SDO-AIA filters, the three jet footpoints have similar morphology during peak flaring times. Using both the C2015 and H2012 inversions, we show that J2 and J6 exhibit multi-thermal plasma distributions, while J3 shows a wider Gaussian-like temperature distribution, centered around hot emission. 

$\bullet$ When studying the geyser, we observed that the pre-flare background conditions are similar for all three eruptions but the individual eruptions have different geometry. The geometrical and thermal parameters are unique, where all eruptions have distinct EM distributions. These recurrent jets are not compatible with a homologous self-repeating eruption scenario.
 
$\bullet$ When studying the jets outflow material, the average temperature distribution profiles show that J3 has a broader temperature distribution while J6 seems to be comprised of distinct multi-thermal plasma threads. The J2 has a less pronounced multi-thermal distribution. Two main strands exist for all three jets; one manifesting in the low temperature intervals and one in the higher temperature range for J2 and J6. These visible strands are morphologically unique and are separated spatially.
 
$\bullet$ The RHESSI X-Ray source reconstruction showed distinct 12 - 25 KeV vs. 6 - 12 KeV X-Ray source morphologies, in all our three cases. This hinted that the 12 - 25 KeV emission is mostly attributed to hard X-Ray emission from impact sites of down-streaming non-thermal electron beams. The 6 - 12 KeV emission is mostly due to soft X-Rays resulting from thermal emission from the heated loop tops.
 
$\bullet$ A spectral fitting of the X-Ray sources was pursued. For a temporal interval corresponding to background conditions all three RHESSI spectra are modeled by a thermal distribution of X-Ray plasma. During each of the three jet's peak times, no single thermal model accurately reproduced any of the three spectra. These flare peaks were approximated by a combination of multi-thermal power models and thick target bremsstrahlung models. The result augments the imaging technique and shows that both thermal emission and non-thermal downwards electron beams exist for all three jets. The X-Ray thermal component was in partial agreement with the SDO-AIA EUV estimates.
 
$\bullet$ Based on consistent results from both SDO-AIA and RHESSI observations, we argue that the recent improvements in the CHIANTI database allows for tighter constraints on inverted plasma DEM at higher temperatures of $\log T_e/K>6.8$.

$\bullet$ The results from the DEM and X-Ray analysis are consistent with a blowout minifilament eruption scenario or with QPPs. Both scenarios involve multiple subsequent flare peaks as were seen in our EUV, inverted EM, and X-Ray timeseries data.

$\bullet$ The solar wind stream is currently discontinuous in regions close to the solar surface. AR jets might offer substantial input in both mass and energy to the solar wind flux. At least in the case of the geyser studied here, AR jets appear to be stronger by roughly two order of magnitude than their polar coronal hole counterparts. The jets of AR11302 escape into the inner heliosphere, but we speculate that although energetic, AR jets lack the ubiquitousness of their polar counterparts, limiting their potential influence on heliospheric energetics and dynamics.

$\bullet$ Can the geyser be considered a typical microflare site? From a thermal perspective the geyser erupts with power around the lower limits of X-Ray thermal microflares. The geyser was found to be also compatible with a non-thermal emitting microflare site. Our data can not distinguish if chromospheric evaporation is the main process that drives non-thermal electron thermalization as envisioned by the standard flare model.

$\bullet$ We show that jet eruptions from penumbral sites  are compatible with basic standard flare model assumptions, and emphasize the importance of the scale independence of reconnection when studying flaring phenomena at different energy classes.

~\\

The authors thank the anonymous reviewer for the very pertinent comments that significantly improved this work. In addition, we thank Drs. Gabriel Dima and Daniela Lacatus for the initial review of this manuscript. A.R.P. and P.G.J. were funded by The  National  Center  for  Atmospheric  Research, sponsored  by  the  National Science Foundation under cooperative agreement No. 1852977. A.R.P is likewise grateful for support through Monash University, The Monash School of Mathematical Sciences, the Astronomical Society of Australia and through an Australian Government Research Training Program (RTP) Scholarship. Raw data and calibration instructions are obtained courtesy of NASA/SDO-HMI, SDO-AIA, and STEREO-EUVI science teams. The authors welcome and appreciate the open data policy of the SDO and STEREO missions. CHIANTI is a collaborative project involving George Mason University, the University of Michigan (USA), University of Cambridge (UK) and NASA Goddard Space Flight Center (USA). This work has made use of NASA's Astrophysics Data System (ADS). 

~\\
 
\bibliography{bibliography} 
 
\appendix

\section{Differential Emission Measures of Thermal Plasma}\label{sec-appendix}
\subsection{On Differential Emission Measure Inversions}\label{sec-appendix:sec-theory}

The Solar EUV emission lines of ions are formed over a wide range of temperature and density regimes. Therefore line ratios and DEM techniques are an essential tool for the determination of plasma conditions. Intensity ratios of density sensitive emission lines or filtergram pairs have become a familiar sight.

DEMs are the solution to an inverse problem resulting from emission line intensities of a partially or fully ionized optically thin astrophysical plasma of constant abundances and in local thermal equilibrium. We follow the work of \citet{craig1986} interpretation of the ill-posed inversion problem where the total line intensity of a two level atom for a given transition $i$ occurring from a upper level $u$ to a lower level $l$ is given by,
     \begin{equation}
I_i~ = \frac{h\nu_i}{4\pi}\int_z n^i_u(z) \;A_{ul}\; dz \qquad [W\;m^{-2}\;sr^{-1}],
  \label{sec-appendix:eq-atom}
  \end{equation}
where $h$ is the Plank constant, $\nu_i$ is the frequency of an emission line, $A_{ul}$ is the Einstein coefficient for spontaneous de-excitation and $n^i_u(z)$ is the population density of the upper level of the $i$ transition. The intensity is assumed to be integrated along the entire line profile.  For a coronal gas in a collisionless approximation, the rate of spontaneous de-excitation ($A_{ul}$) dominates over the collisional de-excitation rate ($C_{ul}$).
 
The population density $n_u^i$ can be approximated by $n_e^2$ multiplied with a function of temperature \citep{craig1986}.
The intensity $I$ is derived from the spectroscopic or filtergram integrated signal in an area cross-section,  
     \begin{equation}
I_i~ = \int_{-z}^{~z} n^{2}_{e}~G_i(T_{e}) ~dz  \qquad [W\;m^{-2}\;sr^{-1}],
  \label{sec-appendix:eq-ifil}
  \end{equation}
  where $G(T_{e})$ represents a response function comprised by the atomic and physical constraints. Note that in practice this is a function of a pre-determined temperature interval. Integrating in depth ($z$) we can recover a volume (usually in cm$^3$), thus enabling us to estimate the average plasma density contained within the volume. A plasma filling factor quantity needs to be assumed.

Eq. \ref{sec-appendix:eq-ifil} can be rewritten in terms of temperature space by considering:
      \begin{equation}
 n^{2}_{e} ~dz  \leftrightarrow DEM(T_{e})~dT_{e},~\qquad\rightarrow~ DEM(T_{e}) ~=~ \frac{n^{2}_{e} ~dz  }{ dT_{e}}   \qquad [ \text{cm}^{-5}\cdot K^{-1}~].
  \label{sec-appendix:eq-dl2dte}
  \end{equation}
 The plasma is fitted in [$T_{e},T_{e}+dT_{e}$] intervals, and is presumed of constant $n_{e}$ density inside each discrete temperature bin \citep{craig1986}. In a standard case, the response function can be simplified by revealing that the DEM represents the distribution of emitting material at a given temperature inside a volume:

     \begin{equation}
I_i~ = ~G_i(T_{e})\int_{T_{e}}^{} DEM(T_{e})~ dT_{e}.
  \label{sec-appendix:eq-ifil2}
  \end{equation}
 
We present the temperature integrated total EM, as opposed to the more traditional DEMs, to compute the total emission of a number of emitting particles from the LOS volume bounded by the cross-sectional area emitting inside a temperature range. We do this in order to recover the total amount of electrons that are heated to different temperatures. In practice, the integrated EM in eq. \ref{sec-appendix:eq-em_th} can be obtained from the observable $I$ and the response function $G_i(T_{e})$ using either of eqs. \ref{sec-appendix:eq-ifil} or \ref{sec-appendix:eq-ifil2}.
\begin{equation}
 EM =\int_{-z}^{~z} n^{2}_{e} ~dz~=~\int_{T_{e}}^{} DEM(T_{e})~ dT_{e}  \qquad [ \text{cm}^{-5}~].
\label{sec-appendix:eq-em_th}
\end{equation}

The SDO-AIA EUV filters, which record optically thin coronal plasma in narrowband wavelengths can not be used to directly infer plasma densities due to their multi-thermal response. Practically, one potential filter ratio will manifest for multiple temperatures across the observed plasma, which has very wide temperature response in the range of $\log T_{e}/K =[5.5,7.5]$.

\begin{equation}
\begin{bmatrix}
I_1\\
I_2\\
...\\
I_n\\
\end{bmatrix}=
\begin{bmatrix}
G_{11} & G_{12} & ...& G_{1n}\\
G_{21} & G_{22} & ...&G_{2n}\\
...&...&...&...\\
G_{m1} & G_{m2} & ...&G_{mn}\\
\end{bmatrix}
\begin{bmatrix}
DEM_1\\
DEM_2\\
...\\
DEM_m\\
\end{bmatrix}.
\label{sec-appendix:eq-em_obs}
\end{equation}

Multiple filtergram observations, with individual temperature sensitivities are coupled inside a system (eq. \ref{sec-appendix:eq-em_obs}), and compared to sets of theoretical responses, with the aim of calculating the best theoretical response solutions that match all $I_{1-n}$. Mathematically, this represents an inverse problem. First, a set of small temperature bins under which the theoretical DEM can be calculated are generated, where the logarithmic temperature range can be divided into uniform linear intervals by considering $T_{bin}=\log T_{e}$.

  \begin{equation}
EM(T_{bin}) ~=~ n^{2}_{e}~\frac{ ~dl  }{ d \log T_{e}}~=~\frac{T_{e}\cdot DEM(T_{e})}{ln(10)} .
\label{sec-appendix:eq-dem2em}
\end{equation}

Eq. \ref{sec-appendix:eq-dem2em} can be used to calculate the total EM inside such one temperature bin from the DEMs following eqs. \ref{sec-appendix:eq-dl2dte}-\ref{sec-appendix:eq-em_th}.

\subsection{The Filter Ratio Technique}\label{sec-appendix:sub-fr}

 The most straightforward approach for plasma EM and density determination is a implementation of the `filter ratio technique' used on Hinode XRT filter inversions \citep[see ][]{paraschiv2015,pucci2013}. This method requires prior knowledge of the temperature of the emitting plasma and assumes that the analyzed emitting volume is isothermal along the line of sight. As previously discussed, eq. \ref{sec-appendix:eq-ifil} presents the general physical interpretation of a filtergram intensity signal. In practice the plasma's emitting  temperature ($T_e$) was assumed following the results of the $T_{e}$ calculation done using the H2012 and C2015 methods. The local plasma density can be inferred via any two EM measurements in filters that have comparable response for a given temperature.  

\begin{equation}
I_{fil} = F_{fil}(T_{e})\cdot\int n^{2}_{e} dz =  F_{fil}(T_{e})\cdot \int DEM(T_e) dT_e   \qquad [DN \cdot s^{-1}\cdot pixel^{-1} ].
  \label{sec-appendix:eq:ifil}
  \end{equation}

When analyzing a particular observation the background emission can be subtracted. The remaining counts correspond to the quantity $I_{loop}-I_{bkg}$, where $I_{loop}$ represents the emission of the studied hot structure and $I_{bkg}$ represents the emission of the background corona. Usually plasma emission is attributed to both local background coronal and  hot flaring components, e.g. $n_{e\,(obs)}=n_{e\,(bkg)}+n_{e\,(loop)}$.  Eq. \ref{sec-appendix:eq-ifil} is updated to
 \begin{equation}
\begin{aligned}
I_{cor}  &= F_{fil}(T_{e})\cdot \int_{-\infty}^{\infty} n^{2}_{e\,(bkg)} dz,  \\
I_{loop}&= F_{fil}(T_{e})\cdot  \left[  \int_{-\infty}^{-z/2} n^{2}_{e\,(bkg)} dz + \int_{-z/2}^{z/2} (n_{e\,(bkg)}+n_{e\,(loop)})^2 dz+ \int_{z/2}^{\infty} n^{2}_{e\,(bkg)} dz \right],
  \label{sec-appendix:cor-loop}
  \end{aligned}
  \end{equation} 
where $z$ is the width of the structure along the line of sight.

We can write down the equation for different individual AIA filter pairs, thus obtaining a $2^{nd}$ order linear system for any filter combination pair, 
\begin{equation}
                \begin{array}{llcc}
(I_{loop}-I_{bkg})_{fil1} &= \; F_{fil1}(T_{e})&\cdot \; z \cdot (n_{e\,(loop)}^2+2\cdot n_{e\,(loop)} \cdot n_{e\,(bkg)}),\\
(I_{loop}-I_{bkg})_{fil2} &= \; F_{fil2}(T_{e})&\cdot \; z \cdot (n_{e\,(loop)}^2+2\cdot n_{e\,(loop)} \cdot n_{e\,(bkg)}),
  \label{sec-appendix:icor-iloop}
  \end{array} 
        \end{equation}
where $F_{fil}(T_e)$ is fixed. This approach proved to be unsuited when applied to AIA filters.

\subsection{Single Gaussian Chi-square Minimization} \label{sec-appendix:sub-teem}    

Current inversion scheme solutions employ more robust mathematical methods in order to recover thermal structure from AIA filtergrams. The Chi-Square minimization \citep{pearson1900,cochran1952} represents the backbone of most of the standard inversions. Although both the H2012 and A2013 codes rely on the same mathematical method for inverting the six AIA channels, they use a set of different implementations, physical and methodological assumptions. 
 
 In the context of the A2013 \citep{aschwanden2013} \footnote{\href{http://www.lmsal.com/~aschwand/software/aia/aia\_dem.html}{http://www.lmsal.com/~aschwand/software/aia/aia\_dem.html}} model, the $\chi^2$ test can be computed pixelwise for the six observable AIA filtergrams intensities ($I_{fil}$) and their correspondent instrument temperature response functions ($F_{fil}$), 
 \begin{equation}
\chi^2=\sum_{i=1}^{k}\frac{(O_{i}-E_{i})^2}{E_{i}}\quad\Longrightarrow\quad \chi^2=\frac{1}{3} \sum_{i=1}^{6}\frac{(\,I_{fil[i]}-F_{fil[i]}\,)^2}{\sigma ^2_{I[i]}}.
  \label{sec-appendix:eq-chisq}
  \end{equation}
The $\sigma_I$ quantity represents the standard uncertainty in the individual AIA filter flux counts. This is the equivalent to the $E_i$ term present in the standard definition of the $\chi^2$ test representing  a set of expected values. The flux was read pixelwise and the exposures are normalized to $DN\cdot s^{-1}$  during the prep procedure. $\sigma_I$ should not be confused with $\sigma_{T_e}$ discussed below. The $1/(i-i_{free})=1/3$ factor comes from the third order degree of freedom of the parameters of the Gaussian response functions.
\begin{equation}
\sigma_{I[i]}=\sqrt{I_{fil[i]}}, \qquad \text{for} \quad  i\in[1,6].
  \label{sec-appendix:eq-chi-sigma}
  \end{equation}

The optimal (lowest $\chi^{2}$) solution is selected from the total combinations of DEM solution space for the EM, $T_e$, and $\sigma_{T_e}$ quantities.  The $T_e$  range was restricted to  $0.5\le T_e \le 15$ MK, due to problems arising from low counts in the bottom range and from `flat' response function behavior at the top range of the  interval. A2013 recommends an even more constrained temperature range compared to our above mentioned $T_e$ range but due to the fact that our jets exhibit generally hot loops for which $T_e\ge10$ MK we adopted a more relaxed restriction on the upper $T_e$ limit while still obtaining reasonable uncertainties.

The recovered plasma parameters should ideally represent an isothermal emission along the LOS of each pixel cross-section. This does not necessarily hold true for a generic coronal plasma. We introduce the $\sigma_{T_e}$ parameter which represents a set of Gaussian temperature widths set to be equally spaced in the $T_e$ range that is taken into account when attributing the theoretical filter responses $F_{fil}$ to the bins in the $\log(T_e)$ range.  Following the A2013 implementation we have used the $\sigma_{T_e}=[0.1,0.2,...,1.0]$ range.  

The $T_e$ quantity can thus be interpreted as an `EM weighted temperature' \citep{aschwanden2013} where the dominant component of the emitting plasma is the base of the determination for a pixel LOS cross-section. The caveat of this estimation is the loss of information on potential secondary (non-dominant) emitting plasma at a different $T_e$ bin that is outside the $\sigma_{T_e}$ uncertainty of the dominating plasma.  The $\sigma_{T_e}$ maps denote the selected temperature width from a set of iterable quantities which is interpreted from a physical point of view as the temperature uncertainty in the EM flux resulted from the minimum obtained $\chi^2$.                  

 The recovered DEM,
\begin{equation}
DEM=\frac{1}{\Delta T} \sum^6_{i=1} \frac{I_{fil[i]}}{F_{fil[i]}}  \qquad   [ \text{cm}^{-5}\cdot K^{-1}],
  \label{sec-appendix:eq-dem}
  \end{equation}
is correspondent to the best fit between the Gaussian modeled theoretical response functions and actual detected flux data. As discussed above, the A2013 inversion returns only the DEM quantity correspondent to the peak $T_e$ that was deduced. In order to recover the entire EM of the region (pixel) we have adopted the \ref{sec-appendix:eq-demem} approximation. This is a simplification of the standard formulation used in literature (e.g. see eq. (5) from \citet{aschwanden2013}), that is sufficient given the large uncertainties involved.
\begin{equation}
EM\;=\;DEM\cdot \Delta T \; =\;DEM\cdot[(T_e\,+\sigma_{T_e}) \,-\,( T_e\,-\sigma_{T_e})]\;=\;DEM \cdot 2\sigma_{T_{e}} \quad   [ \text{cm}^{-5}].
  \label{sec-appendix:eq-demem}
  \end{equation} 
A simple but well established geometrical approximation \citep{susino2013,aschwanden2013b} was adopted, 
\begin{equation}
n_{e-loop}=\phi \cdot \sqrt{\frac{EM}{z}} \qquad    [ \text{cm}^{-3}].
\label{sec-appendix:eq-neapprox}
\end{equation}
The $z$ quantity represents the depth of the emitting region. Assuming a cylindrical geometry the quantity is approximated in practice to the width of the structure. The width can be measured by Gauss fitting the transversal edges of the structure then taking the FWHM parameter of the fit. Afterwards, the quantity is usually averaged across the length of the emitting structure. The loop filling factor $\phi$ represents a fractional amount of volume that is occupied by the emitting plasma.

\subsection{Regularized Chi-square Minimization}\label{sec-appendix:sub-demreg} 

The H2012 \citep{hannah2012}\footnote{\href{http://www.astro.gla.ac.uk/iain/demreg/}{http://www.astro.gla.ac.uk/iain/demreg/}} implementation relies on the same `$\chi^2$ test' as a base, but uses an alternative approach. The authors present a regularization method based on resolving the AIA filter flux $I_{fil}$ information as a linear system of six equations, one for each suitable AIA filter, as a function of the theoretical filter response functions $F_{fil}$ as factors for twelve (unknown) $DEM(T_e)$ quantities of selected temperature intervals/bins. Thus, the H2012 method does not resolve only the peak EM measured around the dominant $T_e$ as the A2013 approximation does. 

We obtain an under-determined system which requires a set of external bounds/conditions for extracting any meaningful solutions.  This system is transferred to matrix form and solved using a Singular Value Decomposition (SVD) scheme. The main benefits are:
\begin{enumerate}[label=(\roman*)]
\item The matrix formulation and subsequent SVD solver is operation-wise linear, making implementations fast.
\item These methods produce faster and provide comparative results to more complex methods, e.g. second order SVD and maximum entropy regularizations. \citet{judge1997} offers a comparison between these different methods and a broad discussion over the limitations on inverting emission line spectra. 
\item Linear error propagation also limits the uncertainty of the solutions producing, in theory, `smoother' results (see \citet{craig1977}, \citet{craig1986}, and H2012 for detailed discussions). 
 \end{enumerate}
 
Now, following H2012, we use the differential form of \ref{sec-appendix:eq:ifil} and introduce the unknown desired quantity $DEM(T_e)$. The term can be inserted into the minimization criterion from eq. \ref{sec-appendix:eq-chisq} and solved as a linear system for the six filters along bins in the selected temperature range,
\begin{equation}
\chi^2= \left[\frac{ (I_{fil}- F_{fil}\cdot DEM(T_e))^2}{\sigma^2_I}\right]\rightarrow\;min, \qquad \text{where}:\qquad{\quad fil=[1,6]\;\; \atop \log T_e=[5.7,7.3]}{}.
  \label{sec-appendix:eq-minimizematrix}
  \end{equation}
We can solve this linear system by imposing an additional condition in the form of Lagrangian multipliers,
\begin{equation}
\chi^2= \left[\frac{ (I_{fil}- F_{fil}\cdot DEM(T_e))^2}{\sigma^2_I}+ \lambda(L\cdot(DEM(T_e)-DEM_0(T_e)))^2\right]\rightarrow\;min,
\label{sec-appendix:eq-minimizeh}
\end{equation}
where $L$ represents a constraint matrix, $\lambda$ is the `regularization parameter', and $DEM_0(T_e)$ represents an initial guess solution resulting from solving eq. \ref{sec-appendix:eq-minimizematrix}. The optimal solution $DEM(T_e)$ with its correspondent regularization parameter $\lambda$ can be obtained analytically following the \citet{hansen1992} generalized $SVD$ solution, see eq. 6 of H2012. It is worth noting that due to background subtraction procedures or large uncertainties that may be present in the $I_{fil}$ or $F_{fil}(T_e)$ quantities, a positive (physical!) solution for $DEM_{T_e}$ can not be guaranteed mathematically. To address this a $\lambda$ regularization parameter is selected only if $DEM_{T_e}>0$. 

For the initial guess, eq. \ref{sec-appendix:eq-minimizeh} is minimized considering $DEM_0(T_e)=0$.  The solution ($DEM_{guess}(T_e)$) obtained is usually `weakly regularized', $\chi^2>10$.  The constraint matrix is then set to $L=I$. 

The second minimization run takes $DEM_0(T_e)=DEM_{guess}(T_e)$ expecting a goodness of fit $\chi^2<2$. The result $DEM(T_e)$ is analytically the result of an unknown differential equation encompassing: heating, cooling, radiative, heat transport, etc. terms. It is normally a differentiable quantity itself and we have set-up $d\,DEM(T_e)=L $ as its first derivative following \citet{kontar2004} and \citet{hannah2012}. Depending on case by case conditions a $DEM(T_e)$ solution can be obtained using higher order $L$ derivatives constraints. This would also shift the weight from the input data in favor of the guesstimate constraints where H2012 state that such a further constraint may be useful in cases of uncertain data measurements. 

 This scheme is applied on a per pixel basis and results are presented as total EM distribution across temperature bins for one time instance over a selected jet region. Loop and jet plasma $T_e$ determinations are done after background corona subtraction via $I-I_0\leftrightarrow EM-EM_0$  leaving only the EM of geyser loop plotted versus plasma temperature. 

\subsection{Sparse Inversion and Minimization} \label{sec-appendix:sub-sparse}

Suppose we arrange the SDO-AIA inversion problem as an under-determined $2^{nd}$ order linear system. The temperature response range can be divided into equally spaced logarithmic intervals $T_{j}+\Delta T_{j}$. Thus eqs. \ref{sec-appendix:eq-ifil2}, and \ref{sec-appendix:eq-em_th} become
     \begin{equation}
I_i~ = ~\sum_{j=0}^{20} \int_{T_{j}}^{T_{j}+\Delta T_{j}} G_i (T_{j}) \cdot DEM(T_{j})~ dT_{j} \rightarrow \; I_i~=~\sum_{j=0}^{20} G_{ij}\cdot EM_{j}.
  \label{sec-appendix:eq-ifil3}
  \end{equation}
$I$ is a vector of six observed SDO-AIA intensities and EM represents a yet unknown solution vector.  Theoretically, $m=6\,;\,0\leq i\leq 5$ accounting for the available SDO-AIA filter observations, and  $n=21\,;\,0\leq j\leq 20$ to account for equally spaced bins along the $\log T_{e}/K = [5.7,7.7]$  range. In practice the temperature is constrained to a smaller domain due to the uncertainty in the SDO-AIA response data. Nonetheless, in the case of SDO-AIA observations $m<n$ holds true. The equation system $I_i=G_{ij}\cdot EM_j$ is severely under-determined in temperature, giving a space of potential EM solutions. 

The C2015 \citep{cheung2015b}\footnote{\href{http://tinyurl.com/aiadem}{http://tinyurl.com/aiadem}} method uses an alternative minimization scheme, not based on the $\chi^2$ principle. The main goal is to converge on a particular solution from the solution set of the above defined system.  As expected, that solution must also best model the underlying physical conditions and assumptions. Sparse solutions have been shown by \citet{candes2006,candes2007} to better reproduce under-determined signal data when compared to $\chi^2$, minimum energy, and least squares minimizations. Formally we can state the approach as
     \begin{equation}
G_{ij}\cdot EM~=~I_{i}, \qquad \text{where:}\qquad EM ~=~\sum_{j=0}^{20}EM_j ~~and~~ EM\rightarrow \;min.
  \label{sec-appendix:eq-cheung}
  \end{equation}

C2015 acknowledge that the solution is not a physically constrained procedurewhere additional constraints are enforced in the minimization of eq. \ref{sec-appendix:eq-cheung}.
     \begin{equation}
~\sum_{j=0}^{20}EM_j \rightarrow min, \qquad \text{constrained~by: \quad} { EM\geq 0~~ \atop  ~~~I-\sigma \leq ~G\cdot EM\leq ~I+\sigma ~~~~\text{if}~ ~I>\sigma}{}.
  \label{sec-appendix:eq-cheung-obs}
  \end{equation}

 In practice the $~\sum_{j=0}^{20}EM_j $ quantity is minimized taking into account that the $EM_j$ solutions need to be positively defined. Also, due to uncertainties in the filtergram intensities $I_i$, the solution space is allowed to slightly deviate from the rigid interpretations given by eqs. \ref{sec-appendix:eq-em_th}, and \ref{sec-appendix:eq-ifil3}. This deviation can be modeled using a simple Gaussian like sigma, similar to eq. \ref{sec-appendix:eq-chi-sigma}. Following C2015, the deviation value is equated to the more complex  SDO-AIA computed uncertainty obtained via Solarsoft's \emph{aia\_bp\_estimate\_error.pro}.  The routine returns statistical errors along with uncertainties resulting from instrumental errors. 
 
 Lower count filters such as AIA-94\AA{} may not always record DNs higher than the $\sigma$ error estimation for regions that are characterized by lower temperature coronal plasma. In such cases the left side of the double inequality constraint from eq. \ref{sec-appendix:eq-cheung-obs} is set to `0' in order to force the positivity of recovered EM solutions. The inversion method is relying on L1 norms, which can be solved by very fast computational schemes like the simplex algorithm.  The solution is characterized by a low number of initial constraints, does not overfit the data by adding additional assumptions, e.g. regularization or parametrization, and naturally forces positive EM. 



\end{document}